\pdfoutput=1

\documentclass[12pt]{article}
\usepackage{jheppub}

\usepackage{amsmath}
\usepackage{amssymb}
\usepackage{graphicx,color,slashed,braket}
\usepackage{bbm}
\usepackage{ifpdf}
\usepackage{comment}
\usepackage{float}
\usepackage{enumitem}
\usepackage{stmaryrd}

\newcommand{\bB}{\mathbf{B}}
\newcommand{\bC}{\mathbb{C}}  
\newcommand{\bQ}{\mathbb{Q}}  \newcommand{\bR}{\mathbb{R}}

 \newcommand{\cK}{\mathcal{K}} \newcommand{\cN}{\mathcal{N}}
\newcommand{\cO}{\mathcal{O}} \newcommand{\cR}{\mathcal{R}} \newcommand{\cW}{\mathcal{W}}

\newcommand{\be}{\begin{equation}} \newcommand{\ee}{\end{equation}}
\newcommand{\ba}{\begin{eqnarray}} \newcommand{\ea}{\end{eqnarray}}
\newcommand{\lp}{\left(}
\newcommand{\rp}{\right)}

\newcommand{\w}{\wedge}

 \newcommand{\bZ}{\mathbb{Z}}

\def\rmi{{\rm i}}   \def\ii{\rmi} \def\xx{{\mathbf{x}}}
\def\nn{{\mathbf{n}}} \def\kk{{\mathbf{k}}} \def\00{{\mathbf{0}}} \def\11{{\mathbf{1}}}
\def\22{{\mathbf{2}}} \def\33{{\mathbf{3}}} \def\ll{{\mathbf{l}}} \def\LL{{\mathbf{L}}}

 \def\Im{\mathop{\rm Im}\nolimits}

\DeclareMathOperator{\rk}{rank} \DeclareMathOperator{\codim}{codim} 

\makeatletter
\newcommand{\xleftrightarrow}[2][]{%
\ext@arrow 0055{\leftrightarrowfill@}{#1}{#2}%
} \def\leftrightarrowfill@{%
\arrowfill@\leftarrow\relbar\rightarrow%
}
\makeatother

\begin{document}

\title{Stabilizing massless fields with fluxes in Landau-Ginzburg models}

\author[a]{Katrin Becker,}
\emailAdd{kbecker@physics.tamu.edu} 
\author[b]{Muthusamy Rajaguru,}
\emailAdd{muthusamy.rajaguru@lehigh.edu}
\author[a]{Anindya Sengupta,}
\emailAdd{anindya.sengupta@tamu.edu} 
\author[c]{Johannes Walcher,}
\emailAdd{walcher@uni-heidelberg.de}
\author[b]{and Timm Wrase}
\emailAdd{timm.wrase@lehigh.edu}

\affiliation[a]{George P.\ and Cynthia Woods Mitchell Institute for Fundamental Physics and
Astronomy \\
Texas A\&M University, College Station, TX 77843, USA}
\affiliation[b]{Department of Physics, Lehigh University, 16 Memorial Drive East, Bethlehem, PA 18018, USA}
\affiliation[c]{Institute for Mathematics and Institute for Theoretical Physics\\
Ruprecht-Karls-Universit\"at Heidelberg, 69120 Heidelberg, Germany\\}


\abstract{

\noindent
Recent work on flux compactifications suggests that the tadpole constraint generically allows only a
limited number of complex structure moduli to become massive, i.e., be stabilized at quadratic order
in the spacetime superpotential. We study the effects of higher-order terms systematically around
the Fermat point in the $1^9$ Landau-Ginzburg model. This model lives at strong coupling and
features no K\"ahler moduli. We show that indeed massless fields can be stabilized in this fashion. We observe that, depending on the flux, this mechanism is more effective when the number of initially massless fields is large. These findings are compatible with both the massless Minkowski conjecture and the tadpole conjecture but are violating the refined version of the tadpole conjecture. Along the way, we complete the classification of integral
flux vectors with small tadpole contribution. Thereby we are closing in on a future complete
understanding of all possible flux configurations in  the $1^9$ Landau-Ginzburg model.

}

\maketitle

\newpage

\section{Introduction}\label{sec:introduction} 

In the context of string model building, moduli stabilization refers to the lifting of flat
directions in the deformation space of string compactifications by symmetry breaking and dynamical
effects. It has been at the forefront of research in string phenomenology for more than two decades.
The influential early work that proposed various promising scenarios and constructions is reviewed,
for example, in \cite{Grana:2005jc, Douglas:2006es, Blumenhagen:2006ci}. Explicit model building has
however been hampered by many computational challenges as well as deep conceptual problems. In
recent years, the swampland program has emerged as a hopeful guiding principle to disentangle these
complications. Reversing the burden of proof, it calls into question the very existence of
low-energy effective theories that would naturally be expected as part of the string landscape, but
have proven difficult to realize in practice. This encompasses 4-dimensional Anti-de Sitter,
Minkowski, and de Sitter vacua with specific conditions on spectrum and interactions. Given the
absence of massless scalar fields in our universe, moduli stabilization remains the greatest current
challenge among all of these. Continued effort as well as the development of new techniques and
approaches are required to tackle this profound problem. 

In this paper we continue our study \cite{Becker:2006ks, Becker:2007ee, Becker:2007dn,
Bardzell:2022jfh, Becker:2022hse, Cremonini:2023suw, Becker:2023rqi} of this problem in an
orientifold of the $1^9$ Landau-Ginzburg model that is mirror dual to a rigid Calabi-Yau manifold.
It describes the compactification on a ``non-geometric" Calabi-Yau manifold with $h^{1,1}=0$. The
absence of K\"ahler moduli makes it an excellent test case in which to study the stabilization of
complex structure moduli in type IIB flux compactifications. The seminal GKP construction
\cite{Giddings:2001yu} described how fluxes stabilize the complex structure moduli. Early explicit
realizations \cite{Giryavets:2003vd, Denef:2004dm, Denef:2005mm} seemed to confirm the expectation
that generic fluxes will stabilize all complex structure moduli. In the last few years this
expectation has been examined more closely and called into question. In the concrete example of the
sextic Calabi-Yau fourfold, it was observed that there is a tension between satisfying the tadpole
constraint and stabilizing all complex structure moduli~\cite{Braun:2020jrx}. This tension has been
formalized in the tadpole conjecture in \cite{Bena:2020xrh}.  

The tadpole conjecture states that the fluxes used to stabilize moduli contribute to the D3-brane
tadpole by an amount that grows in an unacceptable way the more moduli one wishes to stabilize. In
quantitative terms, the conjecture says that the number $n_{\rm stab}$ of moduli that are
stabilized\footnote{We will discuss the precise definition of this notion momentarily.} for a
specific choice of flux, and the contribution $N_{\rm flux}$ of this flux to the D3-brane tadpole
satisfy the constraint\footnote{The first version of this paper did not have the factor of 2 in the equation below and concluded that the $1^9$ Landau-Ginzburg model does not seem to violate the refined tadpole conjecture. This factor of 2 arises from us defining the flux tadpole in equation \eqref{eq:tadpole} in the covering space following~\cite{Becker:2006ks}. The original tadpole conjecture paper~\cite{Bena:2020xrh} has in equation (2.1) a factor of 1/2 in front of the flux contribution, effectively counting the flux contribution in the quotient space.\\ The tadpole conjecture was further studied in non-geometric LG models in \cite{Becker:2024ayh, Rajaguru:2024emw}, where this point is further clarified. We thank Daniel Junghans for bringing this to our attention.}
\begin{equation}
\label{conjecture}
N_{\rm flux} > 2\, \alpha\, n_{\rm stab}\,,
\end{equation}
where the refined tadpole conjecture states that $\alpha=1/3$.
To preserve supersymmetry, all other contributions to the D3-brane tadpole are positive. They can
only be cancelled by the fixed contribution from the orientifold plane. If \eqref{conjecture} is
correct, this implies that it is not possible to stabilize large numbers of moduli using fluxes. 

The tadpole conjecture has been scrutinized extensively in the asymptotics of moduli space
\cite{Bena:2021wyr, Bena:2021qty, Plauschinn:2021hkp, Lust:2021xds, Marchesano:2021gyv,
Grana:2022dfw, Tsagkaris:2022apo, Coudarchet:2023mmm, Braun:2023pzd} and at special points with
discrete symmetries \cite{Lust:2022mhk}. Our work contributes to a better understanding in the deep
interior of moduli space. Related work on the sextic Calabi-Yau fourfold appears in
\cite{Braun:2023edp}.

The quantities $N_{\rm flux}$ and $n_{\rm stab}$ appearing in \eqref{conjecture} are of paramount
interest for the physics of moduli stabilization. The statement however is in principle of purely
Hodge theoretic nature, as pointed out in particular in \cite{Grana:2022dfw,Becker:2022hse}. The
conjecture is therefore amenable to a completely rigorous analysis. Of course, this depends on a
precise definition of the problem, and in particular of the notion of ``stabilization of moduli''.
As pointed out in \cite{Becker:2022hse}, this is more subtle than one might naively expect. On a
first approach, one might be tempted to simply require that there be no massless fields left in the
supersymmetric vacuum. In mathematical terms, this means that the critical point of the
superpotential $W_{\rm flux}$ induced by the flux should be non-degenerate. For the purposes of the
tadpole conjecture, the quantity $n_{\rm stab}$ would then be defined as the number of erstwhile
moduli that have become massive after turning on the flux. Mathematically, this corresponds to the
rank of the Hessian at the critical point, and leads to a stronger version of the tadpole
conjecture.
\begin{equation}
    \label{stronger}
    n_{\rm stab} := \rk \bigl(\partial_I\partial_J W_{\rm flux}\bigr)
    \quad\leadsto\quad
    \text{stronger version of tadpole conjecture}
\end{equation}
Note that we are here (and also in \eqref{weaker} below) being imprecise in the distinction between
AdS and Minkowski vacua. In fact, for geometric compactifications, there are the well-known GKP type
Minkowski vacua with imaginary self-dual (ISD) fluxes \cite{Giddings:2001yu} and related AdS vacua
with ISD fluxes that appear in the KKLT construction \cite{Kachru:2003aw}. For non-geometric
compactifications, fluxes have to be ISD only for Minkowski vacua that we study in this paper. For
AdS vacua fluxes can contribute with either sign to the tadpole cancellation condition  \cite{Becker:2007dn, Ishiguro:2021csu, Bardzell:2022jfh}. The tadpole conjecture therefore seems mute in that case.

From the physical point of view, massless scalars could be tolerated as long as all flat directions
of the potential are lifted, possibly at higher order in the field expansion. Consider, for example,
a massless scalar field $\phi$ subject to a pure $\phi^4$ potential. Such a field will still mediate
long-range forces. However, cosmological solutions in which it rolls at small constant $\dot \phi$
are impossible. Perturbation theory around the vacuum is in principle well-defined. In fact, one
expects radiative corrections to render the field massive at very low energies. Mathematically, this
means that one should merely require that the critical point of the superpotential be isolated, but
allow that it is possibly degenerate. For this weaker version of the tadpole conjecture, one would
define $n_{\rm stab}$ as the co-dimension of the critical locus.
\begin{equation}
    \label{weaker}
    n_{\rm stab} := \codim \bigl\{ \partial_I W_{\rm flux} = 0 \bigr\}
        \quad\leadsto\quad
    \text{weaker version of tadpole conjecture}
\end{equation}
We understand, of course, that the critical locus need not be a smooth manifold. It can also consist
of several components that intersect at the origin. We will see that this might very well be true in
the case at hand. If so, we define $n_{\rm stab}$ as the {\it minimum} co-dimension of all these
components.

The relation between \eqref{stronger} and \eqref{weaker} follows from the inequality
\begin{equation}
\label{inequality}
\rk \bigl(\partial_I\partial_J W_{\rm flux}\bigr) \le\codim \bigl\{ \partial_I W_{\rm flux} = 0\bigr\}\,.
\end{equation}
Namely, \eqref{stronger} requires {\it less} for \eqref{conjecture} to be true than \eqref{weaker}.
It is hence more difficult to disprove, and therefore physically stronger in that sense.\footnote{In
the reverse (mathematical) sense, \eqref{weaker} is stronger since it {\it claims more} than
\eqref{stronger}.} The distinction between the two versions does not appear in the original
literature cited above. This appears to be due, at least in part, to the absence of \emph{any}
discussion of higher-order terms in the context of moduli stabilization. In our view, it is only the
weaker version \eqref{weaker} that, if true, would really jeopardize ``stabilization of complex
structure moduli by fluxes in the sense of GKP etc.'' Some initial considerations of higher-order
terms in the $1^9$ model can be found in \cite{Becker:2022hse}. The main aim of the present work is
to analyze this more systematically, in light of the weaker version of the tadpole conjecture. We
will find that indeed higher-order terms can stabilize some more massless moduli. For computational
reasons, we have not been able to decide whether the critical points first found in
\cite{Becker:2006ks} are degenerate or not. The technique that we develop along the way however is
general. It can also be applied in other contexts.

We anticipate some other features and limitations of our analysis. As in previous works, we will
study the superpotential around the Fermat point in moduli space. This allows for an easy
calculation of the periods as complete power series, and hence the higher-order terms in the
superpotential. The analysis around other points in moduli space is possible, but more complicated.
We will also restrict the axio-dilaton to $\tau = C_0 + \rmi\, e^{-\phi}=e^{\frac{2 \pi
\rm{i}}{3}}$.  
Thus, we are clearly at strong coupling. We can nevertheless perform exact calculations, if we
restrict to ${\mathcal N}=1$ supersymmetric Minkowski vacua. This is because string loop corrections
only enter the K\"ahler potential \cite{Becker:2006ks}, while the critical point condition remains
holomorphic. The absence of K\"ahler moduli in the $1^9$ Landau-Ginzburg model entails that
\emph{if} we were able to stabilize all moduli, we would in fact not only disprove the weaker
version of the tadpole conjecture, but we would immediately produce Minkowksi vacua of string theory
without any flat directions. It is interesting to remark that by itself this would not disprove the
recently proposed Massless Minkowski conjecture \cite{Andriot:2022yyj}. This conjecture states that
any $\cN=1$ supersymmetric vacuum will admit some massless fields. Again, these massless fields do
not have to give rise to true flat directions. If the stronger form of the tadpole conjecture, based
on \eqref{stronger} remains true, it would imply the persistence of massless fields that could
nevertheless be stabilized at higher order.

We will also pursue the {\it classification} of flux configurations that can stabilize (some of) the
moduli at the Fermat point in the $1^9$ model. This question was also first raised in
\cite{Becker:2006ks}. It arises naturally due to the high rank of the supersymmetric flux lattice. A
systematic study was initiated in the recent paper \cite{Becker:2023rqi}. Specifically, it was
explained how to find many linearly independent integral vectors in the flux lattice that have a
small tadpole contribution $N_{\rm flux}$. In particular, this led to a solution of the shortest
vector problem for the $1^9$ model. Concretely, using exhaustive computer searches, it was shown
that there are no quantized flux solutions that contribute less than $N_{\rm flux}=8$ to the tadpole
cancellation condition. Furthermore, the authors presented a large set of flux configurations that
give $N_{\rm flux}=8$. In this paper we now present \emph{all} flux configuration with such a small
contribution to the tadpole. For the $1^9$ orientifold the flux contribution is bounded $N_{f\rm
lux}\leq 12=N_{O3}/2$ \cite{Becker:2006ks}. Given that there are probably no flux configuration with
$8<N_{\rm flux}<12$ and part of the $N_{\rm flux}=12$ flux configurations have already been
classified in \cite{Becker:2023rqi}, this puts a full classification of all flux configuration for
the $1^9$ model within reach.

The outline of the paper is as follows: In section \ref{sec:review} we review the $1^9$
Landau-Ginzburg model and the ingredients of moduli stabilization. In section \ref{sec:Lattice} we
describe what is known about the set of supersymmetric 3-form fluxes in the model. In particular, we
show that the recent paper \cite{Becker:2023rqi} covers almost all flux configurations with 8 non-zero
components in the $\Omega$-basis (defined in section \ref{sec:review}) and $N_{\rm flux} = 8$. We 
complete this list. In section \ref{sec:higherorder} we evaluate order by order higher terms in the 
superpotential and identify the number of massless fields that are stabilized through higher order terms. 
We summarize our findings in section \ref{sec:conclusion}.

\section{Review of the model}
\label{sec:review} 

The mirror dual of a rigid Calabi-Yau threefold, i.e., a CY$_3$ manifold with $h^{2,1}=0$, would have 
$h^{1,1}=0$ and hence does not admit a K\"ahler manifold description. Instead, one can resort to the 
more general class of orbifoldized Landau-Ginzburg models \cite{vafaOrbifoldized}, as first studied 
in the context of moduli stabilization in \cite{Becker:2006ks}. In general, an $\cN=(2,2)$
Landau-Ginzburg model can be attached to any world-sheet superpotential $\cW(\{x_i\})$ that is a 
holomorphic and (weighted-)homogeneous function of a set of chiral fields $\{ x_i \}$. The
worldsheet action is of the form
\begin{equation}
\label{LGaction}
S = \int d^2 z d^4 \theta\, \cK \left(\{ x_i,\bar{x_i}\}\right) + \left(\int d^2 z d^2 \theta\, \cW
\left(\{x_i\}\right) + c.c \right) \,.
\end{equation}
Here, $\cK$ is the (worldsheet) K\"ahler potential. It is conjectured that $\cW$ determines 
$\cK$ uniquely at the IR fixed point of the renormalization group flow
\cite{Vafa:1988uu}.
$\cK$ is therefore not required
for the specification of the model. The superpotential itself is invariant along the flow (up to
wavefunction renormalization). The central charge of the IR CFT is given by $\hat c = \sum_i
(1-w_i)$. Here, the $w_i$ are the $U(1)$ R-charges of the $x_i$. They are normalized such that $\cW$
has charge $2$. To construct a 4-dimensional string background, one requires $\hat c=3$. It is then
possible to orbifold by a subgroup of phase symmetries to project the model onto integral $U(1)$
R-charges. This ensures a spacetime supersymmetric string background. We will deal exclusively with
the simplest such model in this paper. This is the so-called $1^9$ model. It has $9$ chiral fields
$x_1,\ldots,x_9$, and superpotential
\begin{equation}
\label{undeformed}
\mathcal{W}\bigl(\{x_i\}\bigr)=\sum_{i=1}^{9}x_{i}^{3}\,.
\end{equation}
The orbifold is by a $\mathbb{Z}_{3}$ group generated by the following action on the chiral fields:
\begin{equation}
\label{orbifold}
g: x_{i}\mapsto\omega\, x_{i}\,.
\end{equation}
Here, and throughout this paper, $\omega\equiv e^{\frac{2 \pi \ii}{3}}$.

In general, the rings formed by chiral and anti-chiral fields in the left- and right-moving sectors
of the above $\mathcal{N}=(2,2)$ superconformal field theory are analogous to cohomology rings of
Calabi-Yau manifolds, of dimension equal to the central charge. They correspond to left/right Ramond
ground states by spectral flow. Specifically, the $(c,c)$ ring arises from the states in the
untwisted sector of the Hilbert space of the theory, and is given by the invariant part of the
Jacobi ring. In the case at hand this is
\begin{equation}
\label{jacobi}
\cR = \biggl[\frac{\mathbb{C}\left[x_1, \ldots, x_9\right]}{\partial_{x_i} \cW\left(x_1, \ldots,
x_9\right)}\biggr]^{\mathbb{Z}_3}\,.
\end{equation}
As a complex vector space, this ring has dimension $170$. It is spanned by monomials of the form
\begin{equation}
\xx^\kk = x_1^{k_1} \cdot x_2^{k_2} \cdots x_9^{k_9}
\end{equation}
where $\kk = (k_1,\ldots,k_9)$ satisfies $k_i\in \{0,1\}$ for all $i$ and $\sum k_i = 0 \bmod 3$.
The elements with $\sum k_i=3$ are the $84$ monomials $x_i x_j x_k$ with $i \neq j \neq k \neq i$.
They form a basis for the allowed marginal deformations of the superpotential $\cW$.
\begin{equation}
\label{deformation}
\cW\bigl(\{x_i\}\bigr) = \sum_{i=1}^{9}x_{i}^{3} \;\;\longrightarrow\;\; \cW\bigl(\{x_i\}; \{ t^\kk
\} \bigr) = \sum_{i=1}^9 x_i^3 - \sum_{\substack{\kk \\[.05cm]
\sum \!k_i=3}} t^\kk \xx^{\kk}
\end{equation}
The deformation parameters $t^\kk$ are analogous to complex structure moduli of a geometric
compactification. Together with the axio-dilaton $\tau = C_0 + \rmi\, e^{-\phi}$ they give rise to
massless spacetime fields that we wish to stabilize. On the other hand, the K\"ahler moduli are
contained in the $(a,c)$ ring. This ring arises from the twisted sector of the orbifold. The $1^{9}$
model orbifolded as in \eqref{orbifold} has only two non-trivial twisted sectors. Therefore, the
$(a,c)$ ring contains no marginal deformations. In particular, there is no volume modulus. This is
one way to see that the model does not have  an interpretation as a geometric compactification manifold. 

\subsection{Middle-dimensional (co-)homology} 
\label{sec:middledimensional}

Because the $1^9$ model is non-geometric, it is not possible to study Ramond and Neveu-Schwarz
fluxes in the usual fashion in the supergravity approximation. However, the vertex operators
creating the corresponding spacetime fields still exist in the worldsheet theory. Their interactions
with the moduli induce a superpotential completely analogous to the geometric formulation. The
fluxes are also subject to the same quantization and tadpole cancellation conditions. We refer to
\cite{Becker:2006ks} for a rigorous justification of these statements. Here, we only broach some
ideas, and summarize the results. Crucially, to describe the wrapped fluxes, we require an integral
homology basis, and to understand the space-time superpotential and tadpole cancellation, the
pairing with cohomology. Physically, one can think of integral homology in terms of supersymmetric
cycles wrapped by D-branes. In type IIB, the cycles that can be threaded by fluxes are represented
by A-branes. The cycles that support the orientifold planes and carry (the analogues of) the
D3/D7-brane tadpole are represented by B-branes.

The undeformed $1^9$ model is an orbifolded tensor product of $\cN=2$ minimal models with smallest
possible central charge $\hat c = \frac 13$. This has a Landau-Ginzburg representation with a single
chiral field, and superpotential
\begin{equation}
\mathcal{W}=x^3\,.
\end{equation}
The A-branes of this model are represented by contours in the $x$-plane that asymptote to regions in
which $\Im(\cW)=0$ \cite{HoriIqbalVafa}. There are three such contours,
$\left(V_{0},V_{1},V_{2}\right)$, shown in Fig.\ \ref{fig:contours} below. These are not independent
cycles, but satisfy the one relation,
\begin{equation}
\label{relation}
V_{0}+V_{1}+V_{2}=0\,.
\end{equation}
Under the $\bZ_3$ action \eqref{orbifold}, they transform as 
\begin{equation}
\label{orbicycle}
g: V_n \mapsto V_{n+1\bmod 3}\,.
\end{equation}
Somewhat fancily, one can think of the charge lattice $\Lambda$ of A-branes in the minimal model as
fitting into the exact sequence,
\begin{equation}
\label{fancily}
0 \to \bZ \to \bZ^3 \to  \Lambda \to 0
\end{equation}
where the middle $\bZ^3$ is generated by the $V_0$, $V_1$, $V_2$, and $\bZ$ represents the relation
\eqref{relation}.

\begin{figure}[H]
    \centering
    \includegraphics[scale=0.5]{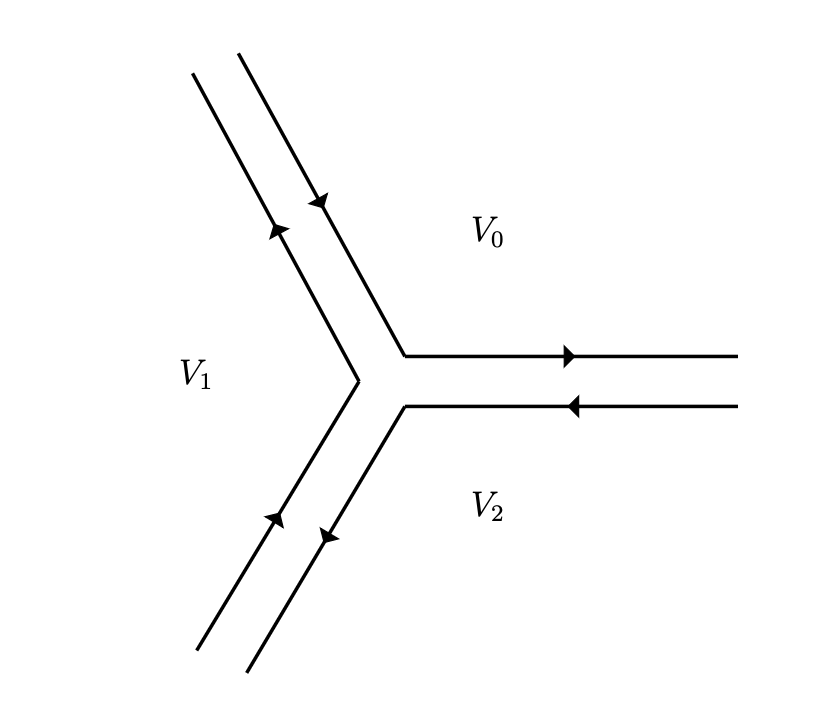}
    \caption{The three contours $\left(V_{0},V_{1},V_{2}\right)$ in the complex $x$-plane.
    \label{fig:contours}}
\end{figure}

The chiral ring of the minimal model is spanned by the elements $1,x\in \cR =\bC[x]/x^2$. These
correspond by spectral flow to Ramond-Ramond ground states traditionally labelled as $\ket{l}$ with
$l=1,2$
\begin{equation}
\label{flow}
x^{k=0,1} \xleftrightarrow{\text{ spectral flow }} \ket{l=1,2}\,.
\end{equation}
The overlap between these Ramond ground states and the boundary states represented by the $V_n$ (the
disk one-point function) can be calculated (after supersymmetric localization) as a contour integral
\cite{HoriIqbalVafa}. Up to normalization, we have
\begin{equation}
\label{singleoverlap}
\braket{V_{n}|l} =
\int_{V_{n}}x^{l-1}e^{-\cW}dx=\frac{1}{3}\omega^{nl}(1-\omega^{l})\Gamma\Bigl(\frac{l}{3}\Bigr)\,,
\end{equation}
where $n\in \{0, 1 , 2 \}$ and $l\in \{1 , 2 \}$. The same integral also calculates the variation of
the overlaps under the deformation $\cW \to x^3 - tx$.
\begin{equation}
\label{defoverlap}
\Bigl(\frac{\partial}{\partial t}\Bigr)^r \Bigr|_{t=0}  \braket{V_n|l} = \int_{V_n} x^{r+l-1}
e^{-x^3} dx = \frac 13 \omega^{n(r+l)}(1-\omega^{r+l})\Gamma\Bigl(\frac {r+l}3\Bigr)\,.
\end{equation}
This vanishes when $r+l=0\bmod 3$ because the integrand is exact in this case. The fact that it does
not vanish when $r+l>2$ (but not $0\bmod 3$), when formally $x^{r+l-1}=0\in\cR$ is zero by the
equations of motion, is physically a result of ``contact terms'' in the operator product expansion.
Mathematically, this amounts to integration by parts. The formula \eqref{defoverlap} will be the
basis for the calculation of the higher-order terms in the superpotential in section
\ref{sec:higherorder}. 

To determine the contribution of the fluxes to the D3-brane tadpole, we require the intersection
form on the charge lattice. Physically, the intersection of $V_{n'}$ and $V_{n}$ can be defined as
the open string Witten index between the respective branes. Mathematically, it is the geometric
intersection between a small counter-clockwise rotation of $V_{n'}$ and $V_{n}$
\cite{HoriIqbalVafa}. In matrix form \cite{Brunner:1999jq},
\begin{equation}
\label{nonsymmetric}
\bigl(\braket{V_{n'}|V_n}\bigr)_{n',n=0,1,2} = \begin{pmatrix} 1 & -1 & 0 \\ 0 & 1 & -1 \\ -1 & 0 &
1 \end{pmatrix} = 1 -g
\end{equation}
where $g$ is the matrix representation of \eqref{orbicycle}. The fact that \eqref{nonsymmetric} is
neither symmetric nor anti-symmetric reflects that a single minimal model is not yet Calabi-Yau.

The calculations are expedited if one uses the Poincar\'e duals of the Ramond ground states as basis
for the charge lattice. This was emphasized in \cite{Becker:2006ks,Becker:2023rqi}. Defining for
$l=1,2$
\begin{equation}
\label{defining}
\Omega_l := \frac 13 \sum_n \omega^{nl} V_n\,,
\end{equation}
with inverse relation 
\begin{equation}
V_n = \sum_l \omega^{-nl} \Omega_l\,,
\end{equation}
we find from \eqref{nonsymmetric}
\begin{equation}
\label{ttstar}
\braket{\Omega_{l'} | \Omega_l} = \delta_{l'+l,3} \,\frac 13 (1-\omega^l)
\end{equation}
and
\begin{equation}
\braket{V_n|\Omega_l} = \frac{1}{3}\omega^{nl}(1-\omega^{l}) \,.
\end{equation}
Thus, by comparison with \eqref{singleoverlap},
\begin{equation}
\ket{l} = \Gamma \Bigl( \frac l3\Bigr) \ket{\Omega_l}\,.
\end{equation}
All these relations are compatible with \eqref{relation} and $\ket{l}=0$ when $l=3$. Eqs.\
\eqref{defining} and the reality of the $V_n$ also imply that complex conjugation acts on the
$\Omega_l$ via 
\begin{equation}
\overline{\Omega_l} = \Omega_{3-l}
\end{equation}
In combination with \eqref{ttstar}, this produces the $tt^*$-metric on the RR ground states
\cite{Cecotti:1991me}.

The full orbifoldized $1^9$ model can now be worked out straightforwardly. The Ramond ground states
are tensor products labelled as $\ket{\ll}$ with $\ll=\left(l_{1},l_{2},\ldots,l_{9}\right)$, $l_{i}
\in \{1 , 2 \}$, and $\sum l_i$ divisible by $3$ in order to satisfy the orbifold projection. These
correspond to the basis of the chiral ring \eqref{jacobi} by spectral flow and can be classified by
Hodge type as shown in table \ref{hodgedeco}.
\begin{table}[t]
\centering
\begin{tabular}{|c|c|c|c|c|}
\hline$\sum_i l_i$ & 9 & 12 & 15 & 18 \\
\hline$H^{(p, q)}$ & $H^{(3,0)}$ & $H^{(2,1)}$ & $H^{(1,2)}$ & $H^{(0,3)}$ \\
\hline
\end{tabular}
\caption{Hodge decomposition of RR ground states in $1^9$ LG model}
\label{hodgedeco}
\end{table}
An (over-complete) integral basis of cycles is obtained by taking tensor products of the $V_n$ to
$V_{\nn}= V_{n_1}\times \cdots \times V_{n_9}$ for $\nn=\left(n_{1},n_{2},\ldots,n_{9} \right)$,
$n_{i} \in \{0, 1 , 2 \}$, and summing over $\bZ_3$ images.
\begin{equation}
\label{account}
\gamma_{\nn} := 
V_{\mathbf{n}}+V_{\mathbf{n+1}}+V_{\mathbf{n+2}}
\end{equation}
where $\mathbf{1}= (1,1,1,1,1,1,1,1,1)$ and $\mathbf{2}=2 \cdot \mathbf{1} = (2,2,2,2,2,2,2,2,2)$.
On the tensor product of \eqref{fancily},
\begin{equation}
0 \to \bZ \to 9 \bZ^3 \to 36 (\bZ^3)^2 \to \cdots \to (\bZ^3)^9 \to \Lambda \to 0 \,,
\end{equation} 
the $\bZ_3$ action is free except on the very first term, where it is trivial. This shows that the
rank of the lattice $\Lambda$ spanned by the $\gamma_{\nn}$ is $((3-1)^9 + 1)/3 -1 = 170$. This is
equal to the dimension of the chiral ring \eqref{jacobi}. The overlap integrals
\eqref{singleoverlap} become
\begin{equation}
\label{generalize}
\braket{\gamma_{\nn}|\ll} = \frac 1{3^8} \omega^{\nn.\ll} \prod_{i=1}^9 (1-\omega^{l_i}) \Gamma 
\Bigl(\frac{l_i}{3}\Bigr)
\end{equation}
where $\mathbf{n}.\mathbf{l}=\sum_{i=1}^9 n_i\, l_i$, and one factor of $3$ is owed to
\eqref{account}. The intersection form is obtained by orbifolding the tensor product \eqref{nonsymmetric}. 
\begin{equation}\label{eq:intersection}
\braket{\gamma_{\nn'}|\gamma_{\nn}} = \braket{V_{\nn'} | V_{\nn}} + 
\braket{V_{\nn'+\11} | V_{\nn}} + \braket{V_{\nn'+\22} | V_{\nn}}
\end{equation}
In the Poincar\'e dual basis 
\begin{equation}
\label{poincdual}
\ket{\Omega_{\ll}} = \frac 1{3^9} \sum_{[\nn]} \omega^{\nn.\ll} \gamma_{\nn} 
= \frac 1{3^9} \sum_{\nn} \omega^{\nn.\ll} V_{\nn} \,,
\end{equation}
the intersection form becomes
\begin{equation}
\label{Omegallbar}
\braket{\Omega_{\ll'}| \Omega_{\ll}} = \delta_{\ll'+\ll,\33}\, \frac 1{3^8} \prod_i (1-\omega^{l_i}) \,.
\end{equation}
We will refer to this as the ``$\Omega$-basis''. Complex conjugation acts on it by
\begin{equation}
\label{comcon}
\overline{\Omega_{\ll}} = \Omega_{\bar \ll} 
\end{equation}
where $\bar\ll=\33-\ll$, and $\33= 3\cdot \11 = (3,3,3,3,3,3,3,3,3)$. The form \eqref{Omegallbar} is
anti-symmetric following the orbifold projection. 

\subsection{Supersymmetric fluxes and tadpole cancellation}
\label{susyfluxes}

We are now in a position to describe supersymmetric 3-form fluxes in the $1^9$ model. There are two
ways to do this. The first is to expand the standard combination of Ramond and Neveu-Schwarz fluxes
$G_3 = F_3 - \tau H_3$ in terms of the integral cohomology basis given by the $\gamma_\nn$. Writing
\begin{equation}
\label{3formbasis} 
G_{3}=\sum_{\mathbf{n}} \left(N^{\mathbf{n}} - \tau M^{\mathbf{n}} \right) \gamma_{\mathbf{n}}\,,
\end{equation}
the $N^{\mathbf{n}}, M^{\mathbf{n}}$ should be integer. They are not uniquely determined because the
$\gamma_{\nn}$ are not linearly independent. The spacetime superpotential induced by this flux is
given by the Landau-Ginzburg version of the standard GVW formula \cite{Gukov:1999ya} 
\begin{equation}
\label{eq:WGVW} W_{{\rm GVW}}= \int \left(F_{3}-\tau H_{3}\right) \wedge \Omega = \braket{G_3|\11}
\end{equation}
Here, we have used table \ref{hodgedeco} to identify the holomorphic three-form $\Omega$ with the
ground state $\ket{\11}$. The overlap should be evaluated with the help of \eqref{generalize}. The
first (and higher) derivatives of the superpotential with respect to the moduli (including the
axio-dilation $\tau$) can be evaluated with the help of \eqref{defoverlap}, see subsection 
\ref{ssec:higherorderderivatives}. Setting them to zero will constrain $G_3$ to be of a certain 
Hodge type as usual. This gives a set of linear equations on the $N^\nn$, $M^\nn$, which have to 
be solved over the integers. The precise formula also depends on the spacetime K\"ahler potential, 
see section \ref{sec:Lattice}.

The alternative approach is to expand $G_3$ in the $\Omega$-basis
\begin{equation}
\label{Omegabasis}
G_3 = \sum_{\ll} A^{\mathbf{l}} \Omega_{\ll}
\end{equation}
This allows to directly constrain its Hodge type by simply setting the undesired $A^\ll$ to $0$.
Flux quantization is equivalent to the condition that in
\begin{equation}
\label{IntegralFluxes} 
\int_{\gamma_{\mathbf{n}}} G_{3} = \braket{\gamma_\nn | G_3} = N_{\mathbf{n}} - \tau
M_{\mathbf{n}}\,,
\end{equation}
which is again to be evaluated with \eqref{generalize}, the $N_{\nn}$ and $M_{\nn}$ have to be integer. 
They are related to the integers in \eqref{3formbasis} by lowering indices with the help of the symplectic 
intersection form \eqref{eq:intersection}.

The two formulations \eqref{3formbasis} and \eqref{Omegabasis} are of course equivalent as far as the 
parametrization of the supersymmetric fluxes is concerned. However, the calculation of the higher-order 
terms in the superpotential is considerably more efficient in the $\Omega$-basis. We therefore prefer it.

The final ingredients are the orientifold projection and the comparison between the O-plane charge
and flux tadpole. These were determined in \cite{Becker:2006ks} using the general formulas provided
in \cite{Hori_2008}. We will restrict to the orientifold of the $1^9$ model that is generated by
dressing worldsheet parity with the exchange of the first two coordinates. This has to be
accompanied by a phase rotation in order to guarantee invariance of the superpotential term in
\eqref{LGaction}. Namely, we are orientifolding by 
\begin{equation}
\label{eq:orientifold} 
\sigma:\left(x_{1},
x_{2}, x_{3}, x_{4}, x_{5}, x_{6}, x_{7}, x_{8}, x_{9}\right) 
\mapsto 
-\left(x_{2}, x_{1},x_{3}, x_{4}, x_{5}, x_{6}, x_{7}, x_{8}, x_{9}\right)\,.
\end{equation}
There are $63$ invariant monomials under this orientifold. Including the axio-dilaton, this gives a
total of $64$ moduli that we wish to stabilize. The above orientifold projection breaks the initial
permutation group $S_9$ of the $1^9$ model to a $\mathbb{Z}_2 \times S_7$ subgroup. We will later
use this group to connect different flux configurations. The O-plane associated with the orientifold
projection in equation \eqref{eq:orientifold} is of ``O3-plane type''. Its charge is equal to $12$
in natural units. This induces a RR tadpole that must be cancelled by the fluxes that we turn on, as
well as possibly adding $N_{\rm D3}$ background D3-branes. The precise condition is that
\begin{equation}
\label{eq:tadpole} N_{\rm flux} = \frac{1}{\tau-\bar{\tau}} \int G_{3} \w \bar{G}_{3} = \int F_3 \w
H_3 \overset{!}{=} 12-N_{\rm D3}\,.
\end{equation}
The overlap of fluxes is to be evaluated with the help of \eqref{Omegallbar} or \eqref{comcon}, if
working with the $\Omega$-basis.

\subsection{The all-order superpotential}
\label{ssec:higherorderderivatives}

By combining \eqref{defoverlap} with \eqref{account}, we obtain the following explicit formula for
an arbitrary multi-derivative of the space-time superpotential \eqref{eq:WGVW} in the $\gamma$-basis 
\begin{equation}
W = \sum (N^\nn - \tau M^\nn\bigl) \braket {\gamma_\nn|\11}
\end{equation}
with respect to the deformation parameters in \eqref{deformation} labelled by $t^{\kk}$, with $\kk$
having nine entries, six of which are $0$ and three of which are $1$. 
\begin{equation}
\label{withrespect}
\frac{\partial}{\partial t^{\mathbf{k}_1}} \frac{\partial}{\partial t^{\mathbf{k}_2}} \ldots 
    \frac{\partial}{\partial t^{\mathbf{k}_r}} 
    \braket{\gamma_\nn|\11}
    \bigg|_{t^{\mathbf{k}}=0} 
    =
    \frac{1}{3^8}
    \omega^{\nn.\LL} 
    \prod_{i=1}^9 
    (1-\omega^{L_i})
    \Gamma\Bigl(\frac{L_i}{3}\Bigr)
\end{equation}
Here, as always, $\omega \equiv e^{\frac{2\pi \rmi}{3}}$, and we have abbreviated
$\LL=(L_1,\ldots,L_9)$ with
\begin{equation}
\label{eq:L}
\LL = \sum_{\alpha=1}^{r} \mathbf{k}_\alpha +{\mathbf{1}}\,.
\end{equation} 
The normalization in \eqref{withrespect} is the same as in \eqref{generalize}. Transforming to the 
$\Omega$-basis
\begin{equation}
W = \sum_\ll A^\ll \braket{\Omega_\ll | \11}
\end{equation}
with the help of \eqref{poincdual}, we find \cite{Becker:2022hse}
\begin{equation}
\label{eq:deriv}
    \frac{\partial}{\partial t^{\mathbf{k}_1}} \frac{\partial}{\partial t^{\mathbf{k}_2}} \ldots 
    \frac{\partial}{\partial t^{\mathbf{k}_r}} \int \Omega_{\bf{l}} \w \Omega\bigg|_{t^{\mathbf{k}}=0} 
    = \delta_{\ll+\LL}\,
    \frac{1}{3^9}\prod_{i=1}^9 
    \lp 1-\omega^{L_{i}}\rp 
    \Gamma\Bigl( \frac{L_{i}}{3}\Bigr)  \,.
\end{equation}
Here, the Kronecker-$\delta$ is understood $\bmod 3$ in all 9 components. Taking account of the
product of $(1-\omega^{L_i})$'s, we find that the derivative in equation \eqref{eq:deriv} vanishes
whenever $L_i=0 \bmod 3$ or $l_i + L_i\neq 0$ mod 3 for any $i\in \{1,2,\ldots, 9\}$. Since all $l_i$ and $(L_i \bmod 3)$ are either $1$ or $2$, the second condition is equivalent to $\bar\ll
= \LL \bmod 3$, where $\mathbf{\bar{l}} = \33-\bf{l}$, and $\33=(3,3,3,3,3,3,3,3,3)$. Because $\ll$ has six
entries equal to $1$ and $3$ entries equal to $2$, we can simplify
\begin{equation}\label{eq:Wder}
    \frac{\partial}{\partial t^{\mathbf{k}_1}} \frac{\partial}{\partial t^{\mathbf{k}_2}} 
    \ldots \frac{\partial}{\partial t^{\mathbf{k}_r}} \int \Omega_{\bf{l}} \w \Omega\bigg|_{t^{\mathbf{k}}=0} 
    =  \begin{cases}
      -\bigl(\sqrt{-3}\bigr)^{-9} 
      \prod_{i=1}^9 \Gamma\bigl(\frac{L_i}{3}\bigr) & \text{for $\bf{\bar{l}}=\bf{L}$ mod 3}\,,\\
       0 & \text{ otherwise}\,. 
    \end{cases}
\end{equation}
Moreover, by the functional equation of the Gamma-function, the product is always a rational
multiple of $\Gamma\bigl(\frac 23\bigr)^6\Gamma\bigl(\frac 13\bigr)^3$. The importance of the result
\eqref{eq:Wder} is computational. It means that before calculating the derivative explicitly, we can
check whether $\mathbf{\bar{l}} =\sum_{\alpha} \mathbf{k}_\alpha + {\bf 1} \bmod 3$. This
substantially speeds up the calculation of higher order terms. We also note that the derivative does
not depend on the individual $\mathbf{k}_\alpha$ but rather only on their sum. 


We now turn to mixed multi-derivatives involving both complex structure moduli and the axio-dilaton.
Since by \eqref{eq:WGVW}, $W$ is linear in $\tau$, we only need to worry about first partial
derivatives with respect to $\tau$. The derivative with respect to $\tau$ can be calculated from
\eqref{3formbasis} and the reality of $F_3$, $H_3$ as usual 
\begin{equation}
\partial_\tau W = \frac{1}{\tau-\bar\tau}\int \bigl(G_3 - \overline G_3\bigr) \wedge \Omega
\end{equation}
In the $\gamma$-basis, this reduces a multi-derivative of the type
\begin{equation}
    \frac{\partial}{\partial \tau} \frac{\partial}{\partial t^{\mathbf{k}_1}} 
    \frac{\partial}{\partial t^{\mathbf{k}_2}} \ldots \frac{\partial}{\partial t^{\mathbf{k}_r}}W 
\end{equation}
to \eqref{withrespect} with the same $\kk_\alpha$'s, but summed only against $M^\nn$'s. In the
$\Omega$-basis, we can use \eqref{comcon} to similarly reduce to \eqref{eq:deriv}. However, we have
to be careful to take into account that in general the coefficients $A^\ll$ will be complex numbers
and also have to be complex conjugated along the way.
For a single complex $A^\ll$ with $G_3=A^\ll \Omega_\ll$ we have
\begin{equation}\label{eq:mixedderivatives}
    \frac{\partial}{\partial \tau} \frac{\partial}{\partial t^{\mathbf{k}_1}} 
    \frac{\partial}{\partial t^{\mathbf{k}_2}} \ldots \frac{\partial}{\partial t^{\mathbf{k}_r}}W 
    =
    \frac{1}{\tau-\bar{\tau}} \int (A^\ll \Omega_{\bf{l}} - \bar{A}^{\ll}\Omega_{\bf{\bar l}}) \w 
    \frac{\partial}{\partial t^{\mathbf{k}_1}} \frac{\partial}{\partial t^{\mathbf{k}_2}} \ldots 
    \frac{\partial}{\partial t^{\mathbf{k}_r}}\Omega \,.
\end{equation}
Using the result \eqref{eq:Wder}, this becomes
\begin{equation}
\label{eq:Wderder}
\frac{\partial}{\partial \tau} \frac{\partial}{\partial t^{\mathbf{k}_1}} 
\frac{\partial}{\partial t^{\mathbf{k}_2}} \ldots \frac{\partial}{\partial t^{\mathbf{k}_r}}W 
\bigg|_{t^{\mathbf{k}}=0, \tau=\tau_0}=  
\begin{cases}
\rmi A^\ll\frac{(\sqrt{-3})^{-9}}{2 \Im(\tau_0)}\,
\prod_{i=1}^9 \Gamma\bigl( \frac{L_i}{3}\bigr) & \text{for $\bar\ll=\LL\bmod 3$} 
\\
\rmi \bar{A}^{\ll}\frac{(\sqrt{-3})^{-9}}{2 \Im(\tau_0)}\,
\prod_{i=1}^9 \Gamma\bigl( \frac{L_i}{3}\bigr) & \text{for $\ll=\LL \bmod 3$}
\\
0 & \text{ otherwise} 
\end{cases} \,,
\end{equation}
where we defined $\tau_0$ to be the vacuum expectation value of the axio-dilaton.
Again, this can be evaluated quite speedily on a computer using only modular arithmetic. Note
however that the contributions from $\ll=\LL$ are proportional to $\Gamma\bigl(\frac
13\bigr)^6\Gamma\bigl(\frac 23\bigr)^3$ and are not rationally related to those from $\bar\ll=\LL$.

By combining all of the above, the exact superpotential for a generic flux $G_3 = \sum_{\ll} A^{\ll}
\Omega_{\bf{l}}$ with complex prefactors $A^{\ll}$  becomes 
\begin{align}
\label{fullglory}
W= -\bigl(\sqrt{-3}\bigr)^{-9} \sum_{\bf{l}} \sum_{r=1}^\infty &\frac{1}{r!}\left(\sum_{\{t^{\mathbf{k}_\alpha}\} \text { with } 
\mathbf{L} = \mathbf{\bar l}} \prod_{i=1}^9 \Gamma\Bigl( \frac{L_{i}}{3}\Bigr) t^{\mathbf{k}_1} 
t^{\mathbf{k}_2} \ldots t^{\mathbf{k}_r}\, A^{\bf{l}} \lp 1-\rmi \frac{\tau-\tau_0}{2 \Im(\tau_0)} \rp\right.\cr
&\left.-\rmi\sum_{\{t^{\mathbf{k}_\alpha}\} \text { with } \mathbf{L} = \mathbf{l}} \prod_{i=1}^9 
\Gamma\Bigl( \frac{L_{i}}{3}\Bigr) t^{\mathbf{k}_1} t^{\mathbf{k}_2} \ldots t^{\mathbf{k}_r}\, 
\bar{A^{\bf{l}}}\, \frac{\tau-\tau_0}{{2 \Im(\tau_0)}}\right)\,.
\end{align}

For the purposes of moduli stabilization, this function has to be restricted to the orientifold
fixed locus $t^\kk = t^{\sigma(\kk)}$. Following \cite{Becker:2023rqi}, we do this in practice by
ordering the $\kk$'s alphabetically and dropping orientifold repetitions. We identify
\begin{equation}
    t^I = t^{\kk_I} = t^{\sigma(\kk_I)}
\end{equation}
with $I\in\{1,\ldots,63\}$ and include the axio-dilaton via
\begin{equation}
 t^0 = \tau-\tau_0\,.
\end{equation}
This gives us finally a flux-dependent and highly transcendental function of $64$ variables whose
critical behaviour at the origin is the subject of the following sections.

\section{The supersymmetric flux lattice}
\label{sec:Lattice}

In most studies of moduli stabilization, one begins with a fixed choice of $3$-form flux $G_3$
within the tadpole bound. The moduli that give rise to vacua preserving $\cN=1$ spacetime
supersymmetry are then solutions of the F-term equations $D_IW=0$. Here, the index $I$ runs over all
moduli including the axio-dilaton, $\tau$. The covariant derivative $D_IW=\partial_IW+ \partial_IK\,
W$ of the Gukov-Vafa-Witten superpotential \eqref{eq:WGVW} depends on the K\"ahler potential $K$. In
geometric compactifications, with the standard dependence of $K\supset -\log\Im(\tau)$, the moduli
have to be adjusted such that $G_3$ is imaginary self-dual (ISD) \cite{Giddings:2001yu}. This
defines a subset in the product of the complex structure moduli space with the upper half-plane that
has been called ``the supersymmetric locus''. The tadpole conjecture \cite{Bena:2020xrh} is
concerned with the co-dimension of this locus, as explained in the introduction. This is a stringent
constraint because, as emphasized in \cite{Denef_2004} the tadpole is {\it positive definite} for
ISD fluxes.

In the setting of the non-geometric $1^9$ Landau-Ginzburg model
\cite{Becker:2006ks,Bardzell:2022jfh,Becker:2022hse}, we describe fluxes that are supersymmetric at
the Fermat point in moduli space.  We also fix the axio-dilaton to a particular value. We call the
set of such fluxes the ``supersymmetric flux lattice''.\footnote{The condition that the flux be
invariant under the orientifold will usually be left implicit.} For any point on this lattice, the
superpotential is critical by definition. We are then interested in the behaviour of the
superpotential around that point, in dependence on the contribution to the D3-brane tadpole. An
important distinction to the geometric situation, emphasized in \cite{Becker:2007dn}, is that the
K\"ahler potential needs to be determined by mirror symmetry. Type IIA string theory compactified on
a rigid $\text{CY}_{3}$ ($h^{2,1}=0$) leads to the K\"ahler potential for the K\"ahler moduli and
axio-dilaton \cite{Grimm:2004ua}
\begin{equation}
\label{KahlerIIA} 
K_{{\rm IIA}} = - 4 \log{[\tau - \bar{\tau}]} - \log{\left[\int_{M}J\wedge J \wedge J\right]}
\,.
\end{equation}
Mirror symmetry exchanges the K\"ahler moduli with complex structure moduli. The K\"ahler potential
is given by 
\begin{equation}
\label{KahlerIIB} 
K = - 4 \log{\left[\tau - \bar{\tau}\right]} -
\log{\left[\int_{M}\Omega \wedge \bar{\Omega}\right]} \,,
\end{equation}
or rather its Landau-Ginzburg analogue, see section \ref{sec:review}. Crucially, this differs by a
factor of $4$ from geometric type IIB compactifications to 4d \cite{Grimm:2004uq}. As a consequence,
the equations $D_IW=0$ do not restrict $G_3$ to be ISD. This was exploited in
\cite{Becker:2007dn,Becker:2007ee,Ishiguro:2021csu,Becker:2022hse}. In this work, we will restrict
to supersymmetric Minkowski vacua. This imposes the additional constraint $W=0$. Then the equations
$D_IW=\partial_I W = 0$ become independent of the K\"ahler potential. They are not affected by
string loop corrections. The candidate instantons that could correct the superpotential are absent \cite{Becker:2006ks, Becker:2007dn, Kim:2022jvv}. The solutions are identical to geometric type IIB compactifications to
Minkowski space, $G_{3} \in H^{2,1}$, except that in non-geometric settings there are no K\"ahler
moduli and it is in principle possible to stabilize \emph{all} moduli with fluxes.

\subsection{An integral basis of the flux lattice}
\label{sec:integral_basis}

We will now write out these conditions in terms of the cohomology basis reviewed in section
\ref{sec:review}. In order to satisfy flux quantization, the coefficients with respect to the
integral basis $\gamma_\nn$ must be integral periods of the torus with complex structure $\tau$. In
order to be supersymmetric, $G_3$ should be purely of Hodge type $(2,1)$. In the dual expansions
\begin{equation}
\label{master}
G_{3}=\sum_\nn \left(N^\nn - \tau M^\nn \right) \gamma_\nn = \sum_\ll A^\ll \Omega_\ll \,.   
\end{equation}
the $N^\nn$, $M^\nn$ are integer, and the $A^\ll$ are zero except when $\sum l_i=12$. The
$\gamma_\nn$ and $\Omega_\ll$ are related by \eqref{poincdual}, \eqref{account}. If this relation
(the ``period matrix'' of the Landau-Ginzburg model) were completely generic, these conditions would
have no non-trivial solution at all. In the situation at hand, in which all period coefficients are
integral linear combinations of $\omega=e^{\frac{2\pi\ii}{3}}$ and $\omega^2$, there are very many.
More precisely, as observed in \cite{Becker:2006ks}, there are still no solutions unless the
axio-dilaton is of the form
\begin{equation}
\label{Dilaton} 
\tau = \frac{a \omega + b}{c \omega + d}\,,
\end{equation}
with integer $a$, $b$, $c$, $d$. This is easiest to see by writing the second condition in
\eqref{master} as $\braket{G_3|\Omega_\ll} = 0$ unless $\sum l_i=15$. For simplicity, we will
restrict to $\tau=\omega$. For this choice, it follows from \eqref{generalize} that one may set all
but one $A^\ll$ in \eqref{master} to zero. Namely, for any $\ll$ with $\sum l_i=12$,
\begin{equation}
\label{singleomega}
G_{(\ll)} = 27(\omega-\omega^2)\Omega_{\ll} 
\end{equation}
(but no smaller multiple of $\Omega_{\ll}$) is an integral flux of type $(2,1)$. Here, and from now on, 
we will replace the subscript `3' on $G$ with labels for various explicit solutions. We will indicate their
physical characteristics by a superscript as they become available. We observe that $G_{(\ll)}$ and
$\omega G_{(\ll)}$ are linearly independent over the integers (in fact, the reals). When $l_1=l_2=1$
or $l_1=l_2=2$, the flux $G_{(\ll)}$ is invariant under the orientifold \eqref{eq:orientifold}. Its
contribution to the D3-brane tadpole is given by \eqref{eq:tadpole} in terms of its length~\eqref{Omegallbar},
\begin{equation}
\frac{1}{\omega-\omega^2} \braket{\overline G_{(\ll)}| G_{(\ll)}} = 27 \,.
\end{equation}
When $l_1\neq l_2$, we need to add the respective orientifold image. The tadpole contribution
doubles. In total, we obtain $126$ linearly independent primitive integral flux vectors
\begin{equation}
\label{nonintegral}
\begin{split}
G_{(\ll,1)}^{[1,27]} = 27 (\omega-\omega^2) \Omega_{\ll} &\qquad
G_{(\ll,2)}^{[1,27]} = 27 (\omega^2-1) \Omega_{\ll} \quad \bigl(l_1=l_2\bigr)
\\
G_{(\ll,1)}^{[2,54]} = 27 (\omega-\omega^2)\bigl(\Omega_{\ll} +\Omega_{\sigma(\ll)} \bigr) & \qquad 
 G_{(\ll,2)}^{[2,54]} = 27 (\omega^2-1) \bigl(\Omega_{\ll} + \Omega_{\sigma(\ll)} \bigr)
\quad \bigl(l_1\neq l_2\bigr)
\end{split}
\end{equation}
The first entry in the superscript square brackets gives the number of non-zero components in the 
$\Omega$-basis, and the second, the tadpole contribution. We also use an additional 
subscript to label different flux choices with the same square bracket superscripts. As a result, the supersymmetric flux lattice in fact has full maximal rank. 

All the fluxes in \eqref{nonintegral} have a tadpole in excess of the orientifold charge (equal to
$12$, see \eqref{eq:tadpole}). Fluxes with smaller tadpole can be constructed by taking suitable
linear (but non-integral!) combinations of \eqref{nonintegral}. For example, one may verify that the 
flux
\begin{equation}
\label{smaller}
\begin{split}
G^{[2,18]}_{(1)} &= 27 
\bigl(\Omega_{1,1,1,1,1,1,2,2,2}- \Omega_{2,2,1,1,1,1,1,1,2}\bigr) 
\\
& =-\frac{1}{3}
\bigl(
G^{[1,27]}_{(\ll_1,1)} +2 G^{[1,27]}_{(\ll_1,2)}
+
G^{[1,27]}_{(\ll_2,1)} +2 G^{[1,27]}_{(\ll_2,2)}
\bigr)
\end{split}
\end{equation}
where $\ll_1=(1,1,1,1,1,1,2,2,2)$ and $\ll_1=(2,2,1,1,1,1,1,1,2)$,
is integral for $\tau=\omega$ and has tadpole $18$ as indicated.  
The flux
\begin{equation}
\label{tad12}
G^{[4,12]}_{(1)} = 9 (\omega-\omega^2) 
\bigl(
-\Omega_{{1,1,1,1,1,1,2,2,2}}
+\Omega_{1,1,1,1,2,1,2,2,1}
+\Omega_{1,1,2,2,1,1,1,1,2}
-\Omega_{1,1,2,2,2,1,1,1,1}
\bigr)
\end{equation}
has tadpole $12$, and 
\begin{equation}
\begin{split}
\label{tad8}
G^{[8,8]}_{(1)} &=
9 \bigl( -\Omega_{1,1,1,2,1,2,1,2,1}+ \Omega_{1,1,1,2,1,2,1,1,2}+
\Omega_{1,1,1,2,1,1,2,2,1} -\Omega_{1,1,1,2,1,1,2,1,2} \\
&\qquad +
\Omega_{1,1,1,1,2,2,1,2,1} - \Omega_{1,1,1,1,2,2,1,1,2} - \Omega_{1,1,1,1,2,1,2,2,1} +
\Omega_{1,1,1,1,2,1,2,1,2} \bigr)
\end{split}
\end{equation}
which was first found in \cite{Becker:2006ks}, has tadpole $8$. The latter two fluxes can hence be 
used to construct $\cN=1$ supersymmetric Minkowski vacua.

To describe the full set of physical flux configurations (in particular, to enumerate
integral flux vectors of tadpole $\le 12$), it is important to first find an integral basis of the
supersymmetric flux lattice consisting of vectors of smallest length possible.\footnote{An integral
basis of a lattice $\Lambda$ is a basis of the vector space $\Lambda\otimes\bQ$ with respect to
which any lattice vector has integral coefficients. Eq.\ \eqref{nonintegral} are not an integral
basis, because (for example) it does not contain \eqref{smaller} in its $\bZ$-span. Finding lattice
vectors of small(est) length in high-dimensional lattices is famously a very hard computational
problem.} This problem was tackled in \cite{Becker:2023rqi}, and solved in a two-step process.
First, a lucky coincidence that we describe momentarily yields an integral basis containing
individual vectors of possibly rather large length. Second, by some judicious computational efforts,
one transforms this into another integral basis with smaller lengths. We do not know whether the 
result is optimal.

The number of $\Omega_\ll$'s of Hodge-type $(2,1)$ is $63$, and the corresponding complex
coefficients $A^\ll$ in equation \eqref{master} parameterize the flux. This amounts to $126$ real
parameters, which we assemble in a vector of $\mathbb R^{126}$. The flux quantization conditions
\eqref{IntegralFluxes}, explicitly
\begin{equation}
\label{recast}
    \braket{\gamma_\nn | G_3} = \frac 1{3^8} \sum_\ll A^\ll \omega^{\nn . \ll} 
    \prod_{i=1}^9 (1-\omega^{l_i})  = N_{\mathbf{n}} - \tau M_{\mathbf{n}}~,
\end{equation}
are linear, complex constraints between the $A^\ll$ and the $340$ integers $N_\nn$, $M_\nn$.
(Actually, only $2\times 128 = 256$ of these are independent because of the orientifold.)
Separating real and imaginary parts, and viewing the set $\cup_\nn \{N_\nn, M_\nn\} = 
\{N_n : n = 1(1)340\}$ as coordinatizing integral points $\bZ^{340}\subset \bR^{340}$, we 
can recast \eqref{recast} in terms of a real linear map\footnote{It is to be noted that equation
\eqref{LatticeRank} are not (real and imaginary parts of) equation \eqref{IntegralFluxes} on the
nose, but a linear transform of it by an invertible $340 \times 340$ matrix.} 
$\mathbf B \in \mathbb R^{340 \times 126}$ from $\bR^{126}$ to $\bR^{340}$ as
\begin{equation}
\label{LatticeRank}
    \sum_{\ll} \mathbf B_{n \ll} A^\ll = N_n\,.
\end{equation}
According to \eqref{nonintegral}, this map hits a lattice of rank $126$ inside $\bZ^{340}\subset \bR^{340}$.
In particular, the matrix $\bB$ has full rank $126$. (This is true on general grounds.) We can pick $126$ 
$\bR$-linearly independent rows from this system. We term the $N_n$'s in the corresponding rows 
\emph{independent} flux quantum numbers, and denote them $\{y_i: i=1(1)126\}$. One can then solve 
the system
\begin{equation}
    \sum_{\ll} \mathbf B_{i \ll} A^\ll = y_i~
\end{equation}
to obtain the $A^\ll$ as linear functions of
$y_i$: $ A^\ll = A^\ll(y_1, \ldots, y_{126})$. Having done this, it is still a non-trivial demand
that the remaining $340 - 126 = 214$ flux numbers are integral. Luckily, this in fact is true, as
the linearly dependent equations in \eqref{LatticeRank} are $\mathbb Z$-linear combinations of the
independent ones. This means that the columns of $[\mathbf B_{i \ll}]^{-1}$ are an integral basis of 
the supersymmetric flux lattice \cite{Becker:2023rqi}. Many of the elements in this basis have large
tadpole values. One would like to swap them for fluxes of smaller length, such as \eqref{tad12}, 
\eqref{tad8} and those presented below. In \cite{Becker:2023rqi}, it was shown that this can be 
done via a convenient ${\it SL}(126,\bZ)$ transformation. This guarantees that the result is still 
an integral basis. See appendix B of \cite{Becker:2023rqi} for the explicit list. 

Having described the rank and an integral basis, we now turn to the problem of finding the finite
set of vectors satisfying the tadpole cancellation condition within the infinite lattice. For a generic 
flux $G_3 = \sum_\ll A^\ll \Omega_\ll$, the contribution
made by each summand to the flux-tadpole is determined completely by its coefficient $A^\ll$ because
of equations \eqref{Omegallbar}, \eqref{comcon}. Doing this in practice, one finds
\cite{Becker:2022hse, Becker:2023rqi} that the contribution to the flux tadpole from each turned-on
$\Omega_\ll$ is a homogeneous quadratic in the $y_i$ with positive integer coefficients and hence positive
integer-valued. Therefore, to catalogue all physical solutions with $N_{\rm flux} \leq 12$, we only
need to turn on at most $12$ of the $\Omega_\ll$'s. This process has been initiated in
\cite{Becker:2023rqi} where the search for physical solutions was organized by the number of
$\Omega_\ll$'s turned on. In the remainder of this section, we summarize some of these results of
\cite{Becker:2023rqi} to get a sense of the flux vectors satisfying the tadpole constraint, and their 
simplest physical characteristics. We also provide some additional details on the classification of 
solutions in this model.

\subsection{Taxonomy of massive moduli}
\label{sec:taxonomy}

By construction, the superpotential \eqref{fullglory} computed in subsection \ref{ssec:higherorderderivatives}
\begin{equation}
    W= \int G_3 \w \Omega = W(t^I)
\end{equation}
as a function of the $64$ moduli remaining after the orientifold and its first derivatives $\partial_IW$ 
vanish at the origin $t^I=0$, for any $G_3$ in the supersymmetric flux lattice described in the previous
subsection. The simplest non-trivial physical invariant is the Hessian,
\begin{equation}
\label{hMM}
 M_{IJ} = \partial_I \partial_J W\,.
\end{equation}
We think of it as the ``holomorphic mass matrix''. As shown in \cite{Bardzell:2022jfh,Becker:2022hse},
its rank gives the number of moduli that are rendered massive by turning on the flux. This is
$n_{\rm stab}$ appearing in the stronger version \eqref{stronger} of the tadpole conjecture.
The result is interesting already for the simplest fluxes listed in \eqref{nonintegral} (which mind you 
are non-physical 
because their tadpole is too large). For the ``$1$-$\Omega$'' fluxes with tadpole $27$, it turns out that 
when $\ll$ has $l_1=l_2=1$, the rank of $M_{IJ}$ is $16$. When $l_1=l_2=2$, it is $22$. For the ``$2$-$\Omega$'' 
fluxes, i.e., $l_1\neq l_2$ (tadpole $54$), it is also $22$. For the record, under the $S_7$ symmetry group,
the $1$-$\Omega$ solutions $G^{[1,27]}_{(\ll,1)}$, $G^{[1,27]}_{(\ll,2)}$ organize into $6$ distinct orbits,
and the $2$-$\Omega$ solutions, in three.

We then proceed by increasing the number of non-zero coefficients in $G_3 = \sum_\ll A^\ll \Omega_\ll$. 
It was found in \cite{Becker:2023rqi} not to be possible to satisfy the tadpole constraint \eqref{eq:tadpole} 
with $2$- or $3$-$\Omega$ fluxes, so we skip the details such as minimum-$N_{\rm flux}$ solutions of these 
types and the ranks of the corresponding mass matrices. The interested reader may consult \cite{Becker:2023rqi}.
With four $\Omega_\ll$'s one can produce physical fluxes satisfying \eqref{eq:tadpole}. The smallest
value of $N_{\rm flux}$ in this class is $12$, and is attained by precisely $54$ distinct $S_7$
orbits of solutions. Representatives from these orbits are given in equations \eqref{eq:4OmFamily1},
\eqref{eq:4OmFamily2}, \eqref{eq:4OmFamily3}.
\begin{subequations}
\label{eq:4OmFamily1}
    \begin{align}
        G_3 &= \lp 
        a_1 \Omega_{1, 1, 1, 1, 1, 1, 2, 2, 2} + 
        a_2 \Omega_{1, 1, 1, 1, 1, 2, 1, 2, 2} +
        a_3 \Omega_{1, 1, 1, 1, 2, 1, 2, 1, 2} + 
        a_4 \Omega_{1, 1, 1, 1, 2, 2, 1, 1, 2}
        \rp \\
        & {\rm with}~~~ (a_1, \ldots, a_4) = 9 (\omega-\omega^2)\omega^p \left\{
                \begin{array}{ll}
                  (-1,1,1,-1)~,\\
                  (1, -\omega, -1, \omega)~, \\
                  (1, -\omega, -\omega, \omega^2)~, \\
                  (-1, \omega, \omega, -\omega^2)~, 
                \end{array} \right.  ~~~ p=0,1,2
    \end{align}
\end{subequations}
all of which have $16$ massive moduli. The solution $G^{[4,12]}_{(1)}$ given in equations \eqref{eq:G(1)4Om} 
and \eqref{eq:G(1)4Omagain} in the next section belongs to the $S_7$ orbit of the first of these with $p=0$.
\begin{subequations}
\label{eq:4OmFamily2}
    \begin{align}
        G_3 &= \lp 
        a_1 \Omega_{1, 1, 1, 1, 1, 1, 2, 2, 2} + 
        a_2 \Omega_{1, 1, 1, 1, 1, 2, 1, 2, 2} +
        a_3 \Omega_{1, 1, 1, 2, 2, 1, 2, 1, 1} + 
        a_4 \Omega_{1, 1, 1, 2, 2, 2, 1, 1, 1}
        \rp  \\
        & {\rm with}~~~ (a_1, \ldots, a_4) = 9(\omega-\omega^2) \omega^p \left\{
                \begin{array}{ll}
                  (-1,1,1,-1)~,\\
                  (-1,1, \omega, -\omega)~, \\
                  (-1, \omega, 1, -\omega)~, \\
                  (-1, \omega, \omega, -\omega^2)~, \\
                  (1, -\omega, -\omega, \omega^2) ~, 
                \end{array} \right.  ~~~ p=0,1,2
    \end{align}
\end{subequations}
all of which have $22$ massive moduli. The solution $G^{[4,12]}_{(2)}$ given in \eqref{eq:G(2)4Om} belongs to 
the $S_7$ orbit of the first of these with $p=0$.
\begin{subequations}
\label{eq:4OmFamily3}
    \begin{align}
        G_3 &= \lp 
        a_1 \Omega_{1, 1, 1, 1, 1, 1, 2, 2, 2} + 
        a_2 \Omega_{1, 1, 1, 1, 1, 2, 1, 2, 2} +
        a_3 \Omega_{2, 2, 1, 1, 1, 1, 2, 1, 1} + 
        a_4 \Omega_{2, 2, 1, 1, 1, 2, 1, 1, 1}
        \rp \\
        & {\rm with}~~~ (a_1, \ldots, a_4) = 9(\omega-\omega^2)\omega^p \left\{
                \begin{array}{ll}
                  (-1,1,1,-1)~,\\
                  (1,-1, -\omega^2, \omega^2)~, \\
                  (-1, 1, \omega, -\omega)~, \\
                  (1, -\omega, -1, \omega)~, \\
                  (-1, \omega, 1, -\omega) ~, \\
                  (1, -\omega, -\omega^2, 1)~, \\
                  (-1, \omega, \omega^2, -1)~, \\
                  (-1, \omega, \omega, -\omega^2)~, \\
                  (1, -\omega, -\omega, \omega^2)~,
                \end{array} \right.  ~~~ p=0,1,2~,
    \end{align}
\end{subequations}
all of which have $26$ massive moduli. The solution $G^{[4,12]}_{(3)}$ given in \eqref{eq:G(3)4Om} belongs 
to the $S_7$ orbit of the first of these with $p=0$.

\begin{figure}[h]
\centering
\includegraphics[width=0.9\textwidth]{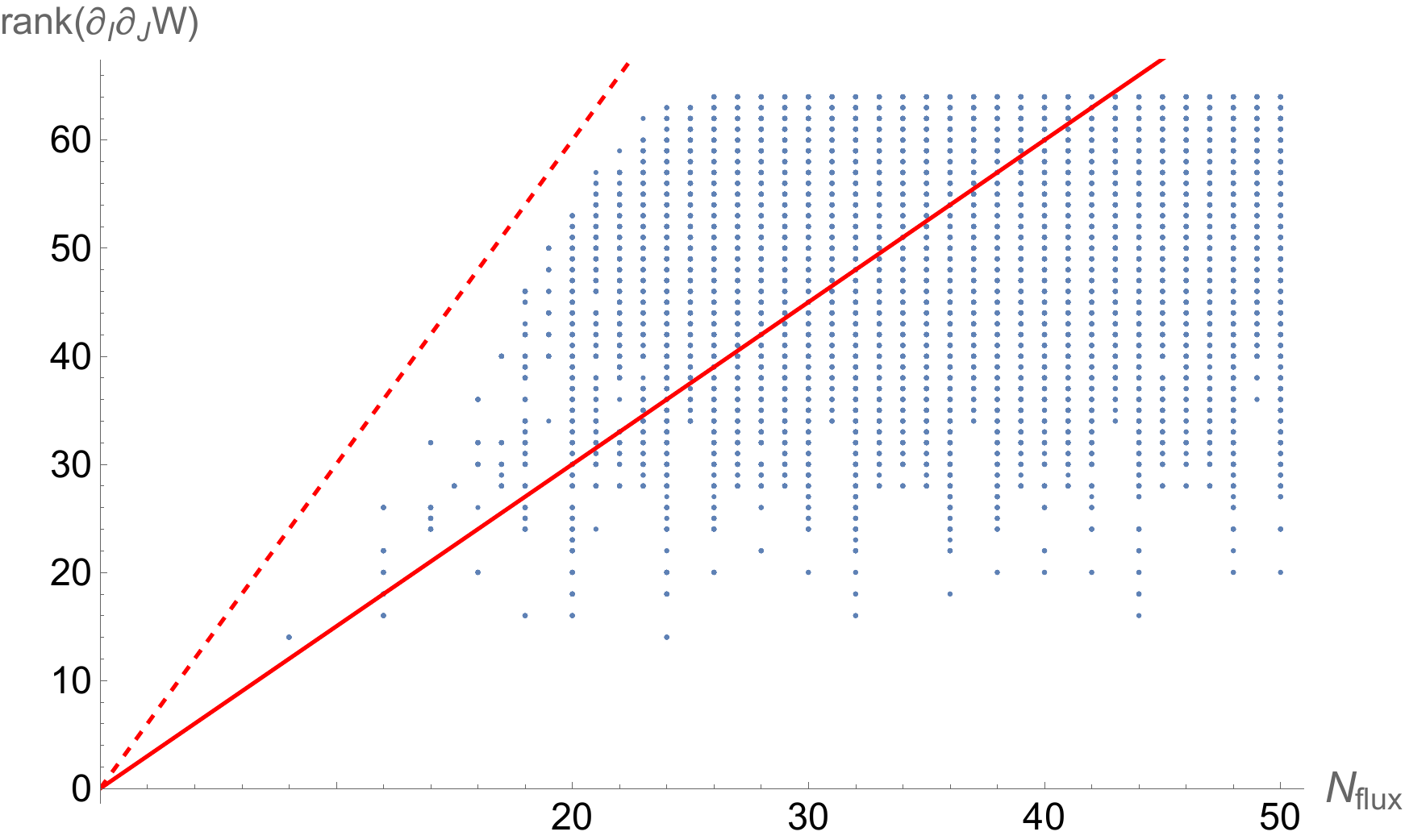}
\caption{A plot of the mass matrix rank vs.\ tadpole contribution for various supersymmetric Minkowski vacua. 
The red solid line with slope 3/2 denotes the bound provided by the refined version of the tadpole conjecture in our conventions. The red dashed line that matches our results well has twice the slope. The total number of moduli in our orientifold of the $1^9$ model is 64. Most flux configurations shown are unphysical since the tadpole cancellation requires $N_{\rm flux} \leq 12$. As described in the text the set of displayed fluxes was generated using particular linear combinations of basis vectors and is only a subset of all possible fluxes for the displayed range of parameters.}
\label{fig:tadvsmass}
\end{figure}

Continuing in this way, it is possible to classify all physical solutions in this model given sufficient 
CPU-hours. We leave this tedious but straightforward task for future work. In anticipation, we have 
generated a large set of fluxes that are small linear combination of the integral basis vectors, and 
evaluated their tadpole and mass matrix rank, by the following process: we generated all linear combinations of up to 4 basis vectors with magnitude one coefficients and for these we computed the tadpole and the mass matrix rank. This is shown in Fig.\ \ref{fig:tadvsmass}, where we plot only results with $N_{\rm flux} \leq 50$. By construction this gives small tadpole fluxes but the set of generated fluxes is, of course, only a subset of all possible fluxes for the displayed range of parameters. 

In our data set, the largest mass matrix rank to $N_{\rm flux}$ ratio is $57/21\sim 2.71$. The corresponding flux is
\begin{equation}
\label{selfproclaimedreigningchampion}
\begin{split}
& G^{[19,21,57]} =
9\bigl(
-\omega \Omega_{{1,1,1,1,1,2,1,2,2}} 
- \Omega_{{1,1,1,1,1,2,2,2,1}} 
-\omega^2 \Omega_{{1,1,1,1,2,1,1,2,2}} 
-\omega^2 \Omega_{{1,1,1,2,1,1,1,2,2}} 
\\ &
+ \omega \Omega_{{1,1,1,2,2,1,1,2,1}} 
+ \omega^2 \Omega_{{1,1,2,1,1,1,2,2,1}} 
+ \omega^2\Omega_{{1,1,2,1,1,2,1,1,2}} 
- \omega \Omega_{{1,1,2,1,1,2,1,2,1}} 
\\ &
+  \Omega_{{1,1,2,1,1,2,2,1,1}} 
+ \omega \Omega_{{1,1,2,1,2,1,1,1,2}} 
+ \omega^2 \Omega_{{1,1,2,2,1,1,1,1,2}}  
- \omega \Omega_{{1,1,2,2,2,1,1,1,1}}  
\\ &
- \omega^2 \Omega_{{1,2,2,1,1,1,2,1,1}} 
+ \omega \Omega_{{1,2,2,1,1,2,1,1,1}} 
- \omega^2 \Omega_{{2,1,2,1,1,1,2,1,1}} 
+ \omega \Omega_{{2,1,2,1,1,2,1,1,1}} 
\\ &
+ \omega^2 \Omega_{{2,2,1,1,1,1,2,1,1}} 
- \omega^2 \Omega_{{2,2,1,1,1,2,1,1,1}}   
-(\omega - \omega^2) \Omega_{{2,2,1,1,2,1,1,1,1}} 
\bigr)
\end{split}
\end{equation}
The above flux and many other data points violate the refined tadpole conjecture in equation \eqref{conjecture}, which required a value smaller than 3/2. However, our data points show a linear relationship as proposed by the tadpole conjecture, albeit with a factor that is closer to 3, see the red dotted line in figure \ref{fig:tadvsmass}. 

We find the smallest tadpole contribution of a flux 
that makes all moduli massive is $N_{\rm flux} = 26$. 
(We do however not know whether this is absolutely the smallest possible.) One such solution is given explicitly as
\begin{equation}
\begin{split}
&G^{[24,26,64]} =
9\bigl(- \omega\Omega_{{1,1,1,1,1,1,2,2,2}} 
+ \omega^2 \Omega_{{1,1,1,1,1,2,1,2,2}} 
- \omega  \Omega_{{1,1,1,1,2,1,2,1,2}}  
\\ &
- \omega^2\Omega_{{1,1,1,1,2,1,2,2,1}}  
-\Omega_{{1,1,1,1,2,2,1,2,1}}  
- \Omega_{{1,1,1,1,2,2,2,1,1}} 
+\omega \Omega_{{1,1,1,2,1,1,2,1,2}} 
\\ &
+\omega \Omega_{{1,1,1,2,2,1,1,1,2}}  
-\omega \Omega_{{1,1,1,2,2,1,1,2,1}} 
+\omega \Omega_{{1,1,2,1,1,1,2,1,2}}  
-\omega^2 \Omega_{{1,1,2,1,1,1,2,2,1}}  
\\ &
-\omega^2 \Omega_{{1,1,2,1,1,2,1,1,2}}  
+\omega^2 \Omega_{{1,1,2,1,1,2,2,1,1}} 
+\omega \Omega_{{1,1,2,2,1,1,1,2,1}}  
-\omega \Omega_{{1,1,2,2,1,1,2,1,1}}  
\\ &
+ \Omega_{{1,2,1,1,2,2,1,1,1}} 
-\omega \Omega_{{1,2,1,2,1,1,1,1,2}}  
+\Omega_{{2,1,1,1,2,2,1,1,1}} 
-\omega \Omega_{{2,1,1,2,1,1,1,1,2}}  
\\ &
+\omega \Omega_{{2,2,1,1,1,1,1,1,2}} 
-\omega^2\Omega_{{2,2,1,1,1,2,1,1,1}}  
+ \omega \Omega_{{2,2,1,2,1,1,1,1,1}} 
- \Omega_{{2,2,1,1,2,1,1,1,1}} 
\\ &
-(\omega-\omega^2) \Omega_{{2,2,2,1,1,1,1,1,1}} \bigr)
\end{split}
\end{equation}

\subsection{Complete classification of the shortest vector solutions}\label{sec:8Omega}

In the recent paper \cite{Becker:2023rqi}, two of the present authors solved the shortest vector
problem for the $1^9$ model. This result was derived using the observation that having exactly $n$
of the coefficients $A^\ll$ non-zero in $G_3 = \sum_\ll A^\ll \Omega_\ll$ results in a crude lower
bound for the flux tadpole: $N_{\rm flux} \geq n$ \cite{Becker:2022hse}. Already in
\cite{Becker:2006ks} the solution $G^{[8, 8]}_{(1)}$, given below in equation \eqref{eq:G1}, was
found to have tadpole 8. By turning on up to 7 $\Omega_{\bf l}$'s, an exhaustive search was launched
for solutions with tadpole smaller or equal to 7. None was found, proving that 8 is the smallest
value of $N_{\rm flux}$ for Minkowski solutions in this model.

To find more solutions that saturate this bound, an Ansatz was made: 
\begin{equation}
    G_3 \propto 
(-\Omega_{\mathbf{l}_{1}}+
\Omega_{\mathbf{l}_{2}}-
\Omega_{\mathbf{l}_{3}}+
\Omega_{\mathbf{l}_{4}}-
\Omega_{\mathbf{l}_{5}}+
\Omega_{\mathbf{l}_{6}}-
\Omega_{\mathbf{l}_{7}}+
\Omega_{\mathbf{l}_{8}})~,
\end{equation}
where the ${\bf l}_a$ vectors are indexed in a certain way (see \cite{Becker:2023rqi} for more
details). The space of 8-$\Omega$ combinations being too large for an exhaustive search, this
simplifying Ansatz was made inspired by the case of 4-$\Omega$ solutions where the solutions
generating the lowest value $N_{\rm flux}=12$ belong to families of the form
\begin{equation}
    G_3 \propto 
(-\Omega_{\mathbf{l}_{1}}+
\Omega_{\mathbf{l}_{2}}-
\Omega_{\mathbf{l}_{3}}+
\Omega_{\mathbf{l}_{4}})~.
\end{equation}
The flux quantization condition, combined with the Ansatz above, implies that 
\begin{equation} 
(-\mathbf{l}_1+\mathbf{l}_{2}-\mathbf{l}_{3}+\mathbf{l}_{4}-
\mathbf{l}_{5}+\mathbf{l}_{6}-\mathbf{l}_{7}+\mathbf{l}_{8})_i\!\!\!\mod 3 =0,
\end{equation}
significantly reducing the number of 8-$\Omega$ combinations allowed. An exhaustive search then
yielded 14 different 8-$\Omega$ solutions with tadpole 8, each having 14 massive moduli. Despite the
success in finding solutions, one finds the lack of proper justification of the Ansatz somewhat
unsatisfactory. Furthermore, this Ansatz was restricted to only non-orientifold fluxes, meaning
$\mathbf{l}_a$ vectors of the kind $(1,1, \ldots)$ and $(2,2, \ldots)$.

Prompted by this, we have now relaxed this Ansatz and made an exhaustive search through all possible
8-$\Omega$ combinations, including cases where one, two, three, or four orientifold fluxes are
turned on. We find that the only choices of $8$ distinct $\Omega_{\mathbf l}$'s that can yield
tadpole 8 are the ones presented in \cite{Becker:2023rqi}. Moreover, all solutions arising from one
of these 8-$\Omega$ choices can be mapped to those from the remaining ones via $S_7$
transformations, which explains why all of the solutions in \cite{Becker:2023rqi} have the same
number of massive moduli. Therefore, it suffices to look for solutions of the form 
\ba
G_3 &=& 9 \lp a_1 \Omega_{1,1,1,1,1,1,2,2,2}+ a_2 \Omega_{1,1,1,1,1,2,1,2,2}+ a_3
\Omega_{1,1,1,1,2,2,1,1,2}+ a_4 \Omega_{1,1,1,1,2,1,2,1,2} + \right.\cr && \quad\,\left.  a_5
\Omega_{1,1,1,2,1,2,1,2,1} + a_6 \Omega_{1,1,1,2,1,1,2,2,1} + a_7 \Omega_{1,1,1,2,2,1,2,1,1} + a_8
\Omega_{1,1,1,2,2,2,1,1,1} \right),\qquad
\ea
with $a_i \in \mathbb{C}$, which belong to the first family presented in section 3.2.3 of
\cite{Becker:2023rqi}, such that the flux is properly quantized and has tadpole 8. One finds a set
of 162 solutions, which can be further modded by the action of the subgroup of $S_7$ that keeps the
choice of the above eight $\mathbf l$ vectors invariant. There are exactly 21 distinct solutions (all of them have $14$ massive moduli) up
to the action of this stability subgroup. These correspond to:
\begin{align}
\label{eq:8Oms}
    (a_1, \ldots, a_8) &= \omega^p \left\{
                \begin{array}{ll}
                  (-1,1,-1,1,-1,1,-1,1) \\
                  (1,-1,\omega,-\omega,1,-1,\omega,-\omega) \\
                  (-1,1,-\omega,\omega,-\omega,\omega,-\omega^2,\omega^2) \\
                  (-1,\omega,-\omega,1,-\omega,1,-1,\omega) \\
                  (1,-\omega,\omega^2,-\omega,\omega,-1,\omega,-\omega^2) \\
                  (-1,\omega,-\omega^2,\omega,-\omega^2,\omega,-\omega^2,1) \\
                  (1,-\omega,\omega^2,-\omega,\omega^2,-\omega,\omega^2,-1)
                \end{array} \right. ~, ~~~ p=0,1,2~.
\end{align}
This leads to the conclusion that, up to the symmetries of the model, there are 21 shortest vectors
in this lattice. The solution $G^{[8, 8]}_{(1)}$, found originally in \cite{Becker:2006ks} and given 
above in equation~\eqref{tad8}, is in the $S_7$ orbit of the first solution in \eqref{eq:8Oms} with $p=0$. 

\section{Moduli stabilization at higher order}
\label{sec:higherorder} 

In this section, we study the stabilization of massless fields via the higher-order terms in the
superpotential that we discussed in subsection \ref{ssec:higherorderderivatives}. We will use
previously studied flux choices and calculate explicitly the higher-order terms and how they
stabilize massless fields.

\subsection{The algorithm and its limitations}

The main idea, sketched in \cite{Becker:2022hse} and discussed in detail in the introduction, is to
address the tadpole conjecture in its weaker form, in which $n_{\rm stab}$ is defined not as the 
number of massive moduli, but as the number of fields whose vacuum expectations values are not 
free parameters, but determined by the field equations, possibly in terms of other fields that
themselves remain massless to all orders in the expansion. Mathematically, this is the difference 
between the Zariski and Krull co-dimension of the critical locus of the superpotential at the 
origin. To explain this concretely, consider a superpotential
\begin{equation}
\label{spacesupo}
W = W(t^I) \in \bC\llbracket t^0,\ldots,t^N\rrbracket
\end{equation}
that is known as a (formal or convergent) power series in the erstwhile moduli $t^I$, including
the axio-dilaton as $I=0$, and assume that $t^I=0$ corresponds to a supersymmetric Minkowksi 
vacuum. This just means that the first non-vanishing term in the expansion of $W$ is the holomorphic
mass matrix from \eqref{hMM}, i.e., we have
\begin{equation}
\label{expansion}
W = \frac 12 M_{IJ} t^I t^J + \frac 16 C_{IJK} t^I t^J t^K + \cdots
\end{equation}
where $C_{IJK}$ is completely symmetric and $\cdots$ denotes higher order terms. In the following,
we will use the shorthand notation $W_r$ for the terms in $W$ that are of order $r$ in the $t^I$.
Thus by definition $W=\sum_{r=2}^\infty W_r$. There are then two key ideas to study the effect of the
$W_r$ for $r>2$ on the vacuum structure for arbitrary numbers of fields.

The first point is to shift the focus from the critical point equations $\partial_I W=0$ as a
geometric locus to the Jacobi ring of the space-time superpotential,
\begin{equation}
\label{spacejacobi}
R = \frac{\bC \llbracket t^0,\ldots,t^N\rrbracket }
{\langle\partial_I W\rangle}
\end{equation}
Mathematically, $R$ is known as the Milnor ring of the function germ defined by $W$. This ring 
is finite-dimensional as a complex vector space precisely if and only if the origin is an
isolated singularity. Physically, \eqref{spacejacobi} contains those physical operators that remain 
non-trivial and independent after imposing the (static) field equations. This is of course just the 
space-time analogue of \eqref{jacobi}. Intuitively, flat directions in $\{\partial_I W=0\}$, say 
one parameterized (possibly non-linearly) by a field $\phi$, will be detected by the infinite 
number of independent operators $\phi$, $\phi^2$, $\phi^3$, \ldots. In principle, the nature of
the critical locus, its decomposition into branches, their singularities, etc.\ is all contained 
in the algebraic properties of the ring $R$ and can be analyzed with standard computer algebra 
packages. In practice however, this is computationally very expensive when the number of moduli
$N$ becomes large, and one wishes to obtain exact statements that depend on terms $W_r$ for 
arbitrary large $r$.

The second idea, then, is to return to a more geometric picture, but proceed order by order in the
field expansion. For example, including terms up to $r=3$ gives us equations of the form
\begin{equation}
\label{quadratic}
M_{IJ} t^J + \frac 12 C_{IJK} t^J t^K = 0\quad \bmod \langle\text{cub.}\rangle
\end{equation}
where $\langle \text{cub.}\rangle$ are elements of $R$ generated by cubic operators.
If $M_{IJ}$ has full rank, these equations have a unique solution in the neighborhood of the 
origin, which is the origin itself. Namely, all moduli have become massive. When $\rk{M_{IJ}}<N+1$ 
is less than maximal, eqs.\ \eqref{quadratic} only allow us to eliminate that many linear
combinations of operators, in terms of the remaining ones. This means that the Zariski dimension, 
intuitively defined as
\begin{equation}
 \dim^Z(\{\partial_IW\}) = \# \frac{\langle\text{lin.}\rangle}{\langle\text{quadr.}\rangle} = 
 N+1 - \rk(M_{IJ})
\end{equation}
where $\langle\text{lin.}\rangle$ and $\langle\text{quadr.}\rangle$ are the elements of $R$
generated by linear and quadratic operators, respectively, remains non-zero. Eliminating these 
linear operators corresponds to ``solving'' $\rk(M_{IJ})$ of the equations \eqref{quadratic}. 
Doing this, and neglecting any cubic terms as indicated, the remaining equations reduce to a 
set of $N+1-\rk(M_{IJ})$ quadratic equations in the same number of independent variables. 
Some of these equations might vanish identically (this happens quite regularly in 
our examples). Moreover, the number of linearly independent quadratic equations might be
larger than the co-dimension of the subspace they cut out.\footnote{A most famous example 
for this phenomenon is the so-called twisted cubic space curve, image of $(s,t)\mapsto 
(x,y,z,w) = (s^3,s^2t,st^2,t^3)$, which is cut out by the three quadrics $xw=yz$, $y^2=xz$, 
$z^2=yw$, but no subset of two of these.} Alternatively, this can be thought of as eliminating 
this many quadratic operators in favor of the independent ones. These statements will be modified 
by the cubic terms in \eqref{quadratic} originating from $W_4$. The linear operators that we eliminated 
in the first step will acquire cubic terms. The intersection of the non-trivial quadrics will also be 
deformed. It might go down in dimension in the process. Finally, some of the equations that vanished 
identically before, might become non-trivial. And so on it goes to higher order.

In practice, we begin by picking a subset of fields $t^{I_a}$, $a=1,\ldots,\rk(M_{IJ})$ that we 
eliminate by solving the respective linear equations originating from $W_2$. These fields could 
appear either quadratically, like $(t^{I_a})^2$ in $W_2$, 
in which case we solve $\partial_{t^{I_a}}W_2=0$ by setting $t^{I_a}=0$ or they only appear linearly like
$t^{I_a} t^{I_b}$ in which case we can solve $\partial_{t^{I_a}} W_2=0$ for $t^{I_b}$ and $\partial_{t^{I_b}} 
W_2=0$ for $t^{I_a}$. If say $t^{I_a}$ appears in another such term, like $t^{I_a}t^{J}$, then really 
$t^J$ is not on the list of $t^{I_a}$'s and we solve $t^{I_b} \sim -t^J$, while $t^J$ remains unstabilized 
at this order. It is appealing to think of the variables $t^{I_a}$ as ``massive fields''. Strictly speaking, 
we can not decide which combination actually acquires a physical mass without knowledge of the K\"ahler 
potential. It was shown in \cite{Bardzell:2022jfh} that the rank of the physical mass matrix is equal 
to $\rk(M_{IJ})$. At the level of counting degrees of freedom, the procedure is completely correct. Although,
there is some arbitrariness in the selection of the $t^{I_a}$.

Solving linear equations for the massive fields does not only work at this order but actually extends to 
all orders. Once we have the linear order solutions $t^{I_a} = t^{I_a}_{1} + \mathcal{O}(t^2)$, where 
$t^{I_a}_1$ are linear polynomials in independent, so-far unstabilized fields found by solving all the 
$\partial_{t^{I_a}}W_2=0$, we can make the Ansatz $t^{I_a}=t^{I_a}_{1}+t^{I_a}_{2}+\mathcal{O}(t^3)$ and 
plug this into $\partial_{t^{I_a}} (W_2 + W_3) = 0 +\mathcal{O}(t^3)$ to get linear equations for the
$t^{I_a}_{2}$. We can solve these linear equations and find quadratic polynomials in unstabilized fields
as solutions for the $t^{I_a}_{2}$. We can proceed like this to higher order and solve only linear 
equations to get $t^{I_a}= \sum_{r=1} t^{I_a}_{r}$, where each of the $t^{I_a}_{r}$ is a polynomial 
of $r$-th power in the $t$'s that are unstabilized at this order. The upshot is that the $t^{I_a}$ can 
easily be solved for and thereby we satisfy all the equations $\partial_{t^{I_a}}W=0$ to arbitrary order 
in $t$ for all the massive fields $t^{I_a}$ in any given example.

We now focus on the fields that are not massive and their corresponding derivatives of the superpotential. 
Concretely, by solving for the massive fields above we have ensured that $\partial_{t^{J}} W_2=0 +
\mathcal{O}(t^2)$ for all $J$ when plugging in $t^{I_a}= \sum_{r=1} t^{I_a}_{r}$. However, at cubic order 
we have to solve $\partial_{J}(W_2 + W_3)=0 +\mathcal{O}(t^3)$. Plugging in $t^{I_a} = \sum_{r=1} t^{I_a}_{r}$ 
solves a subset of these equations, however, generically there remain additional non-trivial quadratic equations
that we  need to solve. Here things get a bit more complicated. Generically, solving quadratic and higher order 
equations is a non-algebraic operation. It typically involves taking square roots of other combinations of fields. 
Luckily, in our examples this never happens. For reasons that can be traced back to the selection rules 
in \eqref{eq:Wder} and \eqref{eq:Wderder}, the relevant polynomials always involve a sufficient number of 
fields that appear only linearly. We can solve for them by merely inverting some of the remaining, 
independent variables. Similar statements hold at higher order in the expansion, at least as far as we have
explored. The only remaining complication is that these solutions might involve different ``branches'' 
that need to be studied independently. For example, equations $xy=xz=0$ give rise to two components with 
different numbers of stabilized fields: $x=0$ or $y=z=0$.

For the purposes of notation, we append the variables that we thereby eliminate to the list of $t^{I_a}$'s,
but include an additional index $r_a$ that indicates at which order we have done so. Thus, $r_a=2$ for 
$a=1,\ldots,\rk(M_{IJ})$, corresponding to the massive fields, $r_a=3$ for those that we eliminate by
solving the independent quadratic equations, etc. Note again that the number of $a$'s with given $r_a$ 
cannot be deduced from the number of independent non-trivial equations that appear at that order alone, 
and moreover will depend on the branch on which we are working. We will not introduce explicit notation to
distinguish these branches, although this is of course essential in practice. The number of fields
that are stabilized up to order $r$ will be denoted $A_r$. Thus, $A_2=\rk(M_{IJ})$, 
$A_3= A_2 + \#\{a \mid r_a=3\}$, etc.

An important observation is that we cannot trivially solve the next higher order by adding quadratic terms to 
the $t^{I_a}$ with $r_a=3$. For example, we usually find linear solutions for the quadratic equations:
$t^{I_a}=t^{I_a}_{1}+\mathcal{O}(t^2)$. We can and have to extend those as $t^{I_a}=t^{I_a}_{1}+t^{I_a}_2
+\mathcal{O}(t^3)$ but when we plug these back into $\partial_{t^{J}}(W_2+W_3+W_4)=0 + \mathcal{O}(t^4)$ then not all $t^{I_a}_{2}$ will actually appear. We can only solve a subset of these equations using the
$t^{I_a}_{2}$ because some quadratic equations have terms like 
$t^{I_a} t^{I_b} = t^{I_a}_1 t^{I_b}_2 + t^{I_a}_2 t^{I_b}_1 + \cO(t^4)$, 
where both $r_a=r_b=3$.
If both
$t^{I_a}_1=t^{I_b}_1=0$
then the corresponding $t^{I_a}_2$, $t^{I_b}_2$ do not appear at all. While this statement seems contrived, 
this actually does happen in some examples. So, at this stage, things become more complicated but we can 
generically solve some of the higher-order equations by fixing higher-order terms in already stabilized fields. 
If there are then still unsolved equations that involve the so far not stabilized fields only, then we stabilize 
some more fields at this order and proceed to the next higher order. 

Given that we start with a finite number of moduli, $N+1$, initially, the procedure must eventually stabilize,
in the sense that there exists an $r_{\rm max}$ such that $A_{r} = A_{\infty}$ for all $r\ge r_{\rm max}$. 
This can happen at different times on different branches, but again there is a maximum order after which the
only effect can be a change of the explicit shape of the branches, but not their dimension. If $A_\infty=N+1$ 
on all branches, this means that all fields have been stabilized. Otherwise, the minimum $A_{\infty}$ over
all branches is what we take as $n_{\rm stab}$ in the weak form \eqref{weaker} of the tadpole conjecture. 
Mathematically, this corresponds to the Krull co-dimension of the critical locus at the origin.

The algorithm that we just described in principle allows to decide how many moduli are stabilized 
by any given flux. It can also be applied in other background models. Some of the phenomena that we 
alluded to however are not easily captured by toy models. So to make the generic discussion more 
concrete and accessible, we work through the details of a particular example up to cubic order in 
the $1^9$ model in the next subsection. Then we summarize the results of some further calculations 
up to order $r=7$. There are two important challenges.

\begin{enumerate}[leftmargin=*, topsep=.2em, parsep=0cm, itemsep=.2em]
\item While the formulas of subsection \ref{ssec:higherorderderivatives} allow us to \emph{in principle} 
calculate all higher order terms, their number grows quickly. At cubic order, we have just from the 63 complex 
structure moduli $63\cdot64\cdot 65/6=43,680$ terms, which is easy to calculate. At septic order, there are 
1,078,897,248 terms and it becomes problematic to calculate and store them when using a normal laptop.
\item When analysing the stabilization of higher order terms we have to solve $\partial_{t^{J}}W=0$. 
When including cubic terms in $W$ we have to solve generically a large number of coupled quadratic equations, 
which is difficult. At higher order, this would then very quickly become an impossible task. However, we 
surprisingly find that the higher order polynomials remain usually relatively simple and we can normally 
solve them without getting square or higher roots. This might be due to the large number of symmetries in 
this model but it would be important to understand this better.
\end{enumerate}

\subsection{A fully worked example}
\label{ssec:fullworked}

In this subsection we discuss a non-trivial example with massless stabilized fields to cubic order in the 
superpotential. The flux, which was first presented in  \cite{Becker:2023rqi} is given by
\begin{equation}\label{eq:G(1)4Om}
G^{[4,12]}_{(1)} =
9 (\omega-\omega^2) \bigl(-
\Omega_{1,1,1,1,1,1,2,2,2}
+\Omega_{1,1,1,1,1,2,1,2,2}
+ \Omega_{1,1,1,1,2,1,2,1,2}
-\Omega_{1,1,1,1,2,2,1,1,2}
\bigr)
\end{equation}
It has $N_{\rm flux}=12$ and it was shown in \cite{Becker:2022hse} that it has 16 massive complex scalars and that 
its cubic terms lead to 10 linearly independent quadratic constraints. Indeed, the quadratic terms in the 
superpotential \eqref{fullglory} being
\begin{equation}
\label{WW2}
\begin{split}
    W_2 = &  A t^0 (t^1 - t^2 - t^6 + t^8) + B \bigl(
    (t^{48}  -t^{49})(t^{52} - t^{55}) + (t^{47}-t^{50})(t^{53}-t^{54}) \\ 
    & + \frac12(t^{33}-t^{34})( t^{58} - t^{61})
    +\frac12(t^{32} - t^{35}) ( t^{59} - t^{60} ) + t^{56} ( t^{38} - t^{40}- t^{44}+ t^{45}) \\
   & + \frac12t^{62} ( t^{23}  - t^{25} - t^{29}  + t^{30} ) 
    + \frac12t^{63} (t^{13}- t^{15} - t^{19}  + t^{20})\bigr)\qquad
\end{split}    
\end{equation}
where 
\begin{equation}
    A = \frac{\omega-\omega^2}{27}\, \Gamma\bigl(\textstyle \frac 13\bigr)^6 \Gamma\bigl(\textstyle\frac 23\bigr)^3\,,
    \qquad\displaystyle
    B= \frac{2}{9}\,\Gamma\bigl(\textstyle \frac 13\bigr)^3 \Gamma\bigl(\textstyle\frac 23\bigr)^6\,.
\end{equation}
We can easily solve $\partial_I W_2=0$ in terms of the sixteen ``massive" fields
\begin{equation}
t^{I_a} \text{ with }\label{eq:massivefields}
I_a \in \{0,1,13,23,34,35,38,49,50,54,55,56,60,61,62,63 \}.
\end{equation} 
This fixes for example $t^0=0 + \mathcal{O}(t^2)$ and $t^1=t^2+t^6-t^8 + \mathcal{O}(t^2)$. The latter equation shows 
that there is an ambiguity in which fields we identify as ``massive". However, as mentioned above, without knowledge 
of the K\"ahler potential this cannot be resolved. Note that we have already been careful in allowing for higher 
order terms in the massive fields that will become important once we go to higher order. Concretely for this 
example the cubic terms in the super potential are
\begin{equation}
\begin{split}
& W_3  = 
C ( \scriptstyle  
t^{31} t^{36} t^{37}-
t^{31} t^{36} t^{39}-
t^{33} t^{38} t^{41}+
t^{34} t^{40} t^{41}-
t^{31} t^{37} t^{42}+
t^{31} t^{39} t^{42}-
t^{32} t^{38} t^{43} +
t^{32} t^{40} t^{43}+
t^{33} t^{41} t^{44} \\[-.3cm] & \scriptstyle \qquad \quad \, \,+
t^{35} t^{43} t^{44}-
t^{34} t^{41} t^{45}-
t^{35} t^{43} t^{45}+
t^{22} t^{36} t^{46}-
t^{24} t^{36} t^{46}+
t^{21} t^{37} t^{46}-
t^{27} t^{37} t^{46}-
t^{21} t^{39} t^{46}+
t^{27} t^{39} t^{46} \\[-.3cm] & \scriptstyle \qquad \quad \, \,-
t^{22} t^{42} t^{46} +
t^{24} t^{42} t^{46}-
t^{28} t^{38} t^{47}+
t^{28} t^{40} t^{47}-
t^{23} t^{43} t^{47}+
t^{25} t^{43} t^{47}-
t^{26} t^{38} t^{48}-
t^{23} t^{41} t^{48}+
t^{29} t^{41} t^{48} \\[-.3cm] & \scriptstyle\qquad \quad \, \, +
t^{26} t^{44} t^{48}+
t^{26} t^{40} t^{49}+
t^{25} t^{41} t^{49}-
t^{30} t^{41} t^{49}-
t^{26} t^{45} t^{49}+
t^{29} t^{43} t^{50}-
t^{30} t^{43} t^{50}+
t^{28} t^{44} t^{50}-
t^{28} t^{45} t^{50} \\[-.3cm] & \scriptstyle\qquad \quad \, \, +
t^{12} t^{36} t^{51}-
t^{14} t^{36} t^{51}+
t^{11} t^{37} t^{51}-
t^{17} t^{37} t^{51}-
t^{11} t^{39} t^{51}+
t^{17} t^{39} t^{51}-
t^{12} t^{42} t^{51}+
t^{14} t^{42} t^{51}+
\,t^{1} t^{46} t^{51}\,\\[-.3cm] & \scriptstyle \qquad \quad \, \,-
\,t^{2} t^{46} t^{51}\,-
\,t^{6} t^{46} t^{51}\,+
\,t^{8} t^{46} t^{51}\,-
t^{18} t^{38} t^{52}+
t^{18} t^{40} t^{52} -
t^{13} t^{43} t^{52}+
t^{15} t^{43} t^{52}-
\,t^{7} t^{47} t^{52}\,+
\,t^{9} t^{47} t^{52}\, \\[-.3cm] & \scriptstyle\qquad \quad \, \, -
t^{16} t^{38} t^{53}-
t^{13} t^{41} t^{53}+
t^{19} t^{41} t^{53}+
t^{16} t^{44} t^{53}-
\,t^{4} t^{48} t^{53}\,+
t^{10} t^{48} t^{53}+
t^{16} t^{40} t^{54}+
t^{15} t^{41} t^{54}-
t^{20} t^{41} t^{54} \\[-.3cm] & \scriptstyle\qquad \quad \, \, -
t^{16} t^{45} t^{54}+
\,t^{4} t^{49} t^{54}\,-
t^{10} t^{49} t^{54}+
t^{19} t^{43} t^{55}-
t^{20} t^{43} t^{55}+
t^{18} t^{44} t^{55}-
t^{18} t^{45} t^{55}+
\,t^{7} t^{50} t^{55}\,-
t^{9} t^{50} t^{55}  \\[-.3cm] & \scriptstyle \qquad\quad \, \,+
\frac{1}{2} t^{12} t^{21} t^{57}-
\frac{1}{2} t^{14} t^{21} t^{57}+
\frac{1}{2} t^{11} t^{22} t^{57}-
\frac{1}{2} t^{17} t^{22} t^{57}-
\frac{1}{2} t^{11} t^{24} t^{57}+
\frac{1}{2} t^{17} t^{24} t^{57}-
\frac{1}{2} t^{12} t^{27} t^{57}+
\frac{1}{2} t^{14} t^{27} t^{57}\\[-.3cm] & \scriptstyle \qquad\quad \, \,+
\frac{1}{2} t^{1} t^{31} t^{57}-
\frac{1}{2} t^{2} t^{31} t^{57}-
\frac{1}{2} t^{6} t^{31} t^{57}+
\frac{1}{2} t^{8} t^{31} t^{57}-
\frac{1}{2} t^{18} t^{23} t^{58}+
\frac{1}{2} t^{18} t^{25} t^{58}-
\frac{1}{2} t^{13} t^{28} t^{58}+
\frac{1}{2} t^{15} t^{28} t^{58} \\[-.3cm] & \scriptstyle \qquad\quad \, \,-
\frac{1}{2} t^{7} t^{32} t^{58}+
\frac{1}{2} t^{9} t^{32} t^{58}-
\frac{1}{2} t^{16} t^{23} t^{59}-
\frac{1}{2} t^{13} t^{26} t^{59}+
\frac{1}{2} t^{19} t^{26} t^{59}+
\frac{1}{2} t^{16} t^{29} t^{59}-
\frac{1}{2} t^{4} t^{33} t^{59}+
\frac{1}{2} t^{10} t^{33} t^{59} \\[-.3cm] & \scriptstyle\qquad\quad \, \, +
\frac{1}{2} t^{16} t^{25} t^{60}+
\frac{1}{2} t^{15} t^{26} t^{60}-
\frac{1}{2} t^{20} t^{26} t^{60}-
\frac{1}{2} t^{16} t^{30} t^{60}+
\frac{1}{2} t^{4} t^{34} t^{60}-
\frac{1}{2} t^{10} t^{34} t^{60} +
\frac{1}{2} t^{19} t^{28} t^{61}-
\frac{1}{2} t^{20} t^{28} t^{61}\\[-.3cm] & \scriptstyle\qquad\quad \, \,+
\frac{1}{2} t^{18} t^{29} t^{61}-
\frac{1}{2} t^{18} t^{30} t^{61}+
\frac{1}{2} t^{7} t^{35} t^{61}-
\frac{1}{2} t^{9} t^{35} t^{61} \\[-.3cm] & \scriptstyle\qquad\quad \, \,
+ (\omega-\omega^2)[-t^{0} t^{48} t^{52}+
t^{0} t^{49} t^{52}-
t^{0} t^{47} t^{53}+
t^{0} t^{50} t^{53}+
t^{0} t^{47} t^{54} -
t^{0} t^{50} t^{54}+
t^{0} t^{48} t^{55}-
t^{0} t^{49} t^{55}\\[-.3cm] & \scriptstyle\qquad\qquad\quad \quad \, \,\,-
t^{0} t^{38} t^{56}+
t^{0} t^{40} t^{56}+
t^{0} t^{44} t^{56}-
t^{0} t^{45} t^{56}-
\frac{1}{2} t^{0} t^{33} t^{58}+
\frac{1}{2} t^{0} t^{34} t^{58}-
\frac{1}{2} t^{0} t^{32} t^{59}+
\frac{1}{2} t^{0} t^{35} t^{59}\\[-.3cm] & \scriptstyle\qquad\qquad\quad \quad \, \,\,+
\frac{1}{2} t^{0} t^{32} t^{60}-
\frac{1}{2} t^{0} t^{35} t^{60} +
\frac{1}{2} t^{0} t^{33} t^{61}-
\frac{1}{2} t^{0} t^{34} t^{61}-
\frac{1}{2} t^{0} t^{23} t^{62}+
\frac{1}{2} t^{0} t^{25} t^{62}+
\frac{1}{2} t^{0} t^{29} t^{62}\\[-.3cm] & \scriptstyle \qquad\qquad\quad \quad \, \,\,-
\frac{1}{2} t^{0} t^{30} t^{62}-
\frac{1}{2} t^{0} t^{13} t^{63}+
\frac{1}{2} t^{0} t^{15} t^{63}+
\frac{1}{2} t^{0} t^{19} t^{63}-
\frac{1}{2} t^{0} t^{20} t^{63} ]\displaystyle )
\end{split}
\end{equation}
where $C = B/3$.

Solving the 16 equations $D_{I_a} (W_2+W_3)=0$ for the 16 massive fields in equation 
\eqref{eq:massivefields}, we find up to this order
\begin{equation}
\begin{split}
\scriptstyle
t^{0}&
\scriptstyle
= (\omega-\omega^2) \bigl(\Gamma\bigl(\frac 23\bigr) / \Gamma\bigl(\frac 13\bigr) \bigr)^3
\bigl( \frac 23 t^{46} t^{51} + \frac 13 t^{31} t^{57}\bigr)
\scriptstyle
\\[-.3cm]
\scriptstyle
t^{1}&
\scriptstyle
=t^{2}+t^{6}-t^{8}
\\[-.3cm]
\scriptstyle
t^{13}&
\scriptstyle
=t^{15}+t^{19}-t^{20}
\\[-.3cm]
\scriptstyle
t^{23}&
\scriptstyle
=t^{25}+t^{29}-t^{30}
\\[-.3cm]
\scriptstyle
t^{34}&
\scriptstyle
=t^{33}-\frac 13 t^{19} t^{28}+\frac 13 t^{20} t^{28}-\frac 13 t^{18} t^{29}+\frac 13 t^{18} t^{30}-\frac 13 t^{7} t^{32}+\frac 13 t^{9} t^{32}
\\[-.3cm]
\scriptstyle
t^{35}
&
\scriptstyle
=t^{32}-\frac 13 t^{16} t^{25}-\frac 13 t^{15} t^{26}+\frac 13 t^{20} t^{26}+\frac 13 t^{16} t^{30}-\frac 13 t^{4} t^{33}+\frac 13 t^{10} t^{33}
\\[-.3cm]
\scriptstyle
t^{38}&
\scriptstyle
=t^{40}+t^{44}-t^{45}
\\[-.3cm]
\scriptstyle
t^{49}&
\scriptstyle
=t^{48}-\frac 13 t^{19} t^{43}+\frac 13 t^{20} t^{43}-\frac 13 t^{18} t^{44}+\frac 13 t^{18} t^{45}-\frac 13 t^{7} t^{47}+\frac 13 t^{9} t^{47}
\\[-.3cm]
\scriptstyle
t^{50}&
\scriptstyle
=t^{47}-\frac 13 t^{16} t^{40}-\frac 13 t^{15} t^{41}+\frac 13 t^{20} t^{41}+\frac 13 t^{16} t^{45}-\frac 13 t^{4} t^{48}+\frac 13 t^{10} t^{48}
\\[-.3cm]
\scriptstyle
t^{54}&
\scriptstyle
=t^{53}-\frac 13 t^{29} t^{43}+\frac 13 t^{30} t^{43}-\frac 13 t^{28} t^{44}+\frac 13 t^{28} t^{45}-\frac 13 t^{7} t^{52}+\frac 13 t^{9} t^{52}
\\[-.3cm]
\scriptstyle
t^{55}&
\scriptstyle
=t^{52}-\frac 13 t^{26} t^{40}-\frac 13 t^{25} t^{41}+\frac 13 t^{30} t^{41}+\frac 13 t^{26} t^{45}-\frac 13 t^{4} t^{53}+\frac 13 t^{10} t^{53}
\\[-.3cm]
\scriptstyle
t^{56}&
\scriptstyle
=\frac 13 t^{33} t^{41}+\frac 13 t^{32} t^{43}+\frac 13 t^{28} t^{47}+\frac 13 t^{26} t^{48}+\frac 13 t^{18} t^{52}+\frac 13 t^{16} t^{53}
\\[-.3cm]
\scriptstyle
t^{60}&
\scriptstyle
=t^{59}-\frac 23 t^{43} t^{44}+\frac 23 t^{43} t^{45}-\frac 13 t^{7} t^{58}+\frac 13 t^{9} t^{58}
\\[-.3cm]
\scriptstyle
t^{61}&
\scriptstyle
=t^{58}-\frac 23 t^{40} t^{41}+\frac 23 t^{41} t^{45}-\frac 13 t^{4} t^{59}+\frac 13 t^{10} t^{59}
\\[-.3cm]
\scriptstyle
t^{62}&
\scriptstyle
=\frac 23 t^{43} t^{47}+\frac 23 t^{41} t^{48}+\frac 13 t^{18} t^{58}+\frac 13 t^{16} t^{59}
\\[-.3cm]
\scriptstyle
t^{63}&
\scriptstyle
=\frac 23 t^{43} t^{52}+\frac 23 t^{41} t^{53}+\frac 13 t^{28} t^{58}+\frac 13 t^{26} t^{59}
\end{split}
\end{equation}
However when looking at all 64 equations $D_I(W_2+W_3)=0$ we find the additional ten linearly independent quadratic relations
\begin{equation}
\begin{split}
\scriptstyle
 (t^{37}- t^{39}) t^{51}+\frac 12 (t^{22} - t^{24} ) t^{57} = 0
\\[-.3cm] \scriptstyle
( t^{36} - t^{42} ) t^{51}+\frac 12 (t^{21}- t^{27}) t^{57} = 0
\\[-.3cm] \scriptstyle
( t^{37} - t^{39}) t^{46}+\frac 12 (t^{12} - t^{14}) t^{57} = 0
\\[-.3cm] \scriptstyle
( t^{36} - t^{42}) t^{46}+\frac 12 (t^{11} - t^{17}) t^{57} = 0
\\[-.3cm] \scriptstyle
 (t^{36} - t^{42}) (t^{37}- t^{39}) = 0
\\[-.3cm] \scriptstyle
 t^{31}( t^{37}- t^{39}) + (t^{22} - t^{24}) t^{46}+ (t^{12} - t^{14}) t^{51} = 0
\\[-.3cm] \scriptstyle
 t^{31} (t^{36}-  t^{42} )+ (t^{21} - t^{27}) t^{46}+ (t^{11}- t^{17}) t^{51} = 0
\\[-.3cm] \scriptstyle
( t^{22} - t^{24}) (t^{36}-t^{42})+ (t^{21} - t^{27}) (t^{37}-  t^{39} ) = 0
\\[-.3cm] \scriptstyle
( t^{12}- t^{14})( t^{36}-t^{42})+ (t^{11}- t^{17}) (t^{37}- t^{39} ) = 0
\\[-.3cm] \scriptstyle
 (t^{12} - t^{14}) (t^{21}-t^{27)}+ (t^{11}- t^{17}) (t^{22}-  t^{24} )  = 0
\end{split}
\end{equation}
One can calculate the Groebner basis for the above set of polynomial equations and finds the Krull co-dimension 
of the ideal to be 6. This means that 6 additional massive fields get stabilized. However, at higher order or 
in more complicated examples below, it becomes computationally too expensive to do such an analysis and therefore
we quickly review how the above equations can be solved explicitly. This leads to different branches or components 
as briefly mentioned above. Concretely, there are components along which 6 fields are stabilized and others along 
which seven fields are fixed: Let us look at the fifth equation, solve it, plug the solution in the other equations 
and keep solving. For ease of presentation we present only two of the different branches that arise:
\ba
&& 1)\,\, t^{36}=t^{42}  \,\, \leadsto \,\, t^{57}=0\,, t^{37}=t^{39} \,\, \leadsto \,\, t^{46}=0\,, t^{11}=t^{17}\,, t^{12}=t^{14}\\
&& 2)\,\, t^{36}=t^{42} \,\, \leadsto \,\, t^{57}=0 \,\, \leadsto \,\, t^{46}=t^{51}=0 \,\, \leadsto \,\, t^{31}=0\,, t^{11}=t^{17}\,, t^{21}=t^{27}\qquad
\ea
The branch 1) fixes only six fields and the branch 2) fixes seven fields. Given that all branches fix either six or 
seven fields, one might then be tempted to conclude that there are six stabilized fields and one could discard the 
other branches with 7 fixed fields. However, there are two reasons to keep track of all different components of 
solutions: Firstly, it is possible that at higher order the branch with  less stabilized fields suddenly stabilizes 
more fields than another branch. We do not find an explicit example of this below. Secondly, it is possible that we 
cannot pursue the branch with the lowest number of stabilized fields to higher order because it is too complicated. 
If one is able to pursue another branch to higher order, then this other branch provides an upper bound on the number of fields that can get stabilized to higher order. We do find an instance of that were we cannot pursue the branch 
with the smallest number of stabilized fields beyond cubic order but we can pursue another branch up to $W_6$, 
thereby providing a useful upper bound on the maximal number of stabilized fields.

\subsection{More examples and results}

Here we will carry out the above-described procedure for many more explicit examples to higher order 
and summarize the results. The first example was called $G_1$ in \cite{Becker:2022hse} and appeared 
already above in equation \eqref{tad8} but we repeat it here for convenience
\ba
\label{eq:G1}
G^{[8, 8]}_{(1)} &=& 9 \lp -\Omega_{1,1,1,2,1,2,1,2,1}+ \Omega_{1,1,1,2,1,2,1,1,2}+
\Omega_{1,1,1,2,1,1,2,2,1} -\Omega_{1,1,1,2,1,1,2,1,2} \right.\cr && \quad\,\left.  +
\Omega_{1,1,1,1,2,2,1,2,1} - \Omega_{1,1,1,1,2,2,1,1,2} - \Omega_{1,1,1,1,2,1,2,2,1} +
\Omega_{1,1,1,1,2,1,2,1,2} \right)\,.\quad
\ea
The above solution has $14$ massive fields and as was observed in \cite{Becker:2022hse}, there are no further quadratic constraints. So, even when including $W_3$
there are no further stabilized fields at this order, although at this order all 63 complex
structure moduli and the axio-dilaton do appear in the superpotential. Surprisingly, we find that
the same holds true when including $W_4, W_5, W_6$. So, even when including sextic terms in the
superpotential we can solve the 64 equations $\partial_{t^{J}}( W_2 + W_3 + W_4 + W_5 +
W_6) = 0 +\mathcal{O}(t^6)$ purely in terms of the 14 massive fields $t^{I_a}$ 
of $r_a=2$. Given this one might 
wonder whether one can prove that no further stabilization is possible at all. Clearly, this cannot 
result from the unstabilized moduli not being present in $W$ since they already do all appear in 
$W_3$ as remarked above. In general,
one can prove that all moduli always appear the latest in $W_5$ (see appendix \ref{app:allmoduli}).
Thus, one would have to find a more elaborate proof that shows that (some) flat directions remain
because (some) unstabilized fields appear only in a very particular way combined with the massive
fields so that $\partial_{t^k}W=0$ to all orders. We did not succeed with this and it is possible
that some higher orders are non-zero. We thus leave this as a challenge for the future to study this
flux choice $G^{[8, 8]}_{(1)}$ to higher order.

Before discussing flux choices that lead to the stabilization of massless fields via higher order
terms in $W$, we list more examples (see \cite{Becker:2022hse, Becker:2023rqi}) that exhibit the
same behavior as $G^{[8, 8]}_{(1)}$ above. The flux choice
\begin{equation}
    \begin{split}
G^{[12, 12]}_{(1)}&=9 \bigl(
-\Omega_{1,1,1,2,2,2,1,1,1}-\Omega_{1,1,1,2,2,1,2,1,1}-\Omega_{1,1,1,2,2,1,1,2,1} +
\Omega_{1,1,1,2,1,2,1,1,2} \\ 
&\quad + \Omega_{1,1,1,2,1,1,2,1,2} + \Omega_{1,1,1,2,1,1,1,2,2} -
\Omega_{1,1,1,1,2,2,2,1,1} -\Omega_{1,1,1,1,2,2,1,2,1}\cr 
&\quad -\Omega_{1,1,1,1,2,1,2,2,1} +
\Omega_{1,1,1,1,1,2,2,1,2} + \Omega_{1,1,1,1,1,2,1,2,2} + \Omega_{1,1,1,1,1,1,2,2,2}
\bigr)\,,
\end{split}
\end{equation}
leads to $N_{{\rm flux}} = 12$ and $22$ massive fields. The flux choices
\begin{equation}
\begin{split}
G^{[12, 12]}_{(2)} &= 9 \bigl( \Omega_{1,1,1,2,2,1,1,2,1} -
\Omega_{1,1,2,1,2,1,1,2,1} - \Omega_{1,1,2,2,1,1,1,2,1} + \Omega_{1,1,2,2,2,1,1,1,1} \\ 
&\quad -\Omega_{1,2,1,2,2,1,1,1,1} -\Omega_{2,1,1,2,2,1,1,1,1} + \Omega_{1,2,2,1,1,1,1,2,1} +
\Omega_{2,1,2,1,1,1,1,2,1} \\ 
&\quad -\Omega_{2,2,1,1,1,1,1,2,1} +\Omega_{2,2,1,1,2,1,1,1,1} +
\Omega_{2,2,1,2,1,1,1,1,1} -\Omega_{2,2,2,1,1,1,1,1,1} \bigr)\,,
\end{split}
\end{equation}
\begin{equation}
\begin{split}
G^{[12, 12]}_{(3)} &=9\omega^2 
\bigl( \Omega_{1,1,2,2,2,1,1,1,1} -
\Omega_{1,2,1,2,2,1,1,1,1} -\Omega_{2,1,1,2,2,1,1,1,1} + \Omega_{2,2,1,1,2,1,1,1,1} \\
& \quad +\Omega_{2,2,1,2,1,1,1,1,1} - \Omega_{2,2,2,1,1,1,1,1,1} +\Omega_{1,1,1,2,2,2,1,1,1}
-\Omega_{1,1,2,1,2,2,1,1,1} \\ 
& \quad - \Omega_{1,1,2,2,1,2,1,1,1} + \Omega_{1,2,2,1,1,2,1,1,1} +
\Omega_{2,1,2,1,1,2,1,1,1} - \Omega_{2,2,1,1,1,2,1,1,1} \bigr)\,, 
\end{split}
\end{equation}
have both $N_{\rm flux} = 12$ and 26 massive fields \cite{Becker:2022hse}. 

Two new solutions were presented in \cite{Becker:2023rqi} that are given by
\ba
\label{eq:G(2)4Om}
G^{[4, 12]}_{(2)}\! &=& 9 \rmi \sqrt{3} 
\left(-\Omega_{1,1,1,1,1,1,2,2,2}+\Omega_{1,1,1,1,2,1,2,2,1}+\Omega_{1,1,2,2,1,1,1,1,2}-\Omega_{1,1,2,2,2,1,1,1,1}\right),\qquad
\quad\\
\label{eq:G(3)4Om}
G^{[4, 12]}_{(3)}\!&=&9 \rmi \sqrt{3} 
\left(-\Omega_{1,1,1,1,1,1,2,2,2}+\Omega_{1,1,2,1,1,1,2,2,1}+\Omega_{2,2,1,1,1,1,1,1,2}-\Omega_{2,2,2,1,1,1,1,1,1}\right).
\ea
These solutions $G^{[4, 12]}_{(2)}$ and $G^{[4, 12]}_{(3)}$ have both $N_{\rm flux} = 12$ and 22 or 26 
massive fields, respectively. 

For all the solutions $G^{[8, 8]}_{(1)}$, $G^{[12, 12]}_{(1)}$, $G^{[12, 12]}_{(2)}$, $G^{[12,12]}_{(3)}$, 
$G^{[4,12]}_{(2)}$, $G^{[4,12]}_{(3)}$ above we find that no massless fields are being stabilized even when 
including up to sextic terms in the superpotential $W$. 

Now let us look at the more interesting and complicated example
\begin{equation}
\label{eq:G(1)4Omagain}
G^{[4, 12]}_{(1)} =9 \rmi \sqrt{3} \bigl(
\Omega_{1,1,1,1,2,1,2,1,2}-\Omega_{1,1,1,1,2,1,2,2,1} -\Omega_{1,1,1,1,2,2,1,1,2} +
\Omega_{1,1,1,1,2,2,1,2,1}\bigr)\,.
\end{equation}
This example was discussed in detail in the previous subsection \ref{ssec:fullworked} up to cubic order in $W$. 
As we have seen there, we have to solve 10 linearly independent quadratic equations and can do so without generating 
square roots since none of the variables appear quadratically. There are different components and they fix either 
six or seven massless fields. We can now pursue the different components to higher order, keeping in mind that the 
total number of stabilized fields is the smallest number of fixed fields, which is six in this case up to this order. 
Since we do not know how many more fields will get stabilized at higher it is worthwhile to keep track of all components. 

Concretely, there is one component where we have 16 massive fields plus six stabilized massless fields that allow us to solve the higher order constraints up to $\partial_{t^{J}}( W_2 + W_3 + W_4 + W_5 + W_6) = 0 +\mathcal{O}(t^6)$. So, for this component, we find no further stabilized fields and have a total of 22 stabilized fields, 16 of which are massive and 6 of which are massless. There are other components where in addition to the 16 massive fields there  are $6+4+0+0$, $7+1+0+0$ and $7+4+0+0$ fixed massless fields. The smallest number of fixed fields is the number of stabilized fields which turns out to be 22 up to order $t^6$ in $W$. We have summarized this in table  \ref{tab:stabilization}, where a question mark means that we have not been able to solve the corresponding polynomial equations. Note that the $G^{[4, 12]}_{(1)}$ model with $N_{\rm flux}=12$ was not violating the refined tadpole conjecture because it had only 16 massive fields, which is smaller than $12\cdot3/2=18$. However, upon including higher order stabilization all models in table \ref{tab:stabilization} violate the refined version of the tadpole conjecture.

\begin{table}
\begin{center}
\begin{tabular}{|c|c|c|c|c|c|
} 
 \hline
Model & massive & 3rd power & 4th power & 5th power & 6th power 
\\ [0.5ex] 
 \hline
 $G^{[8, 8]}_{(1)}$ & 14 & 0 & 0 & 0 & 0 
 \\ 
 \hline
 $G^{[12, 12]}_{(1)}$ & 22 & 0 & 0 & 0 & 0 
 \\
 \hline
 $G^{[12, 12]}_{(2)}$ & 26 & 0 & 0 & 0 & 0 
 \\
 \hline
 $G^{[12, 12]}_{(3)}$ & 26 & 0 & 0 & 0 & 0 
 \\ 
 \hline
 $G^{[4,12]}_{(2)}$ & 22 & 0 & 0 & 0 & 0 
 \\ 
 \hline
 $G^{[4,12]}_{(3)}$ & 26 & 0 & 0 & 0 & 0 
 \\ 
  \hline
 $G^{[4,12]}_{(1)}$ & 16 & 6 & 0 & 0 & 0 
 \\
 & 16 & 6 & 0 & 0 & ? 
 \\
 & 16 & 6 & 4 & 0 & 0 
 \\
 & 16 & 7 & 1 & 0 & 0 
 \\
 & 16 & 7 & 4 & 0 & 0 
 \\
 \hline
  $G^{[12, 12]}_{(4)}$ & 20 & 2 & 0 & 4 & 1 
 \\ 
 & 20 & 2 & 0 & 0 & 0 
 \\
 \hline
 $G^{[12, 12]}_{(5)}$ & 18 & 2 & ? & ? & ? 
 \\ 
 & 18 & 4 & 0 & 0 & 0 
 \\ 
 \hline
\end{tabular}
\end{center}
\caption{A summary of the different models that we have analyzed. The superscript $[n, N_{\rm flux}]$
on the model denotes the number of $\Omega_{\bf{l}}$
components and the tadpole contribution $N_{\rm flux}$. The subscript labels different flux configurations 
with the same $[n, N_{\rm flux}]$. The second column lists the massive fields and the other columns list the 
number of fields that get fixed due to terms in the superpotential that are polynomials of the $r$-th power 
in the moduli. For some models we find different components that either fix the same or different
numbers of fields, as indicated in the multiple rows for the same model. \label{tab:stabilization}}
\end{table}

There are two previously discussed solutions \cite{Becker:2022hse} for which we calculated the higher order 
terms and for which we also find that there are stabilized but massless fields. For 
\begin{equation}
    \begin{split}
G^{[12, 12]}_{(4)} &= 9 \bigl[- \Omega_{1,1,1,1,2,1,2,1,2} + \Omega_{1,1,1,1,2,1,2,2,1}
+ \Omega_{1,1,1,1,2,2,1,1,2} - \Omega_{1,1,1,1,2,2,1,2,1}	
\\ & \quad + \omega(
-\Omega_{1,2,1,1,1,1,2,1,2} + \Omega_{1,2,1,1,1,1,2,2,1} + \Omega_{1,2,1,1,1,2,1,1,2}
-\Omega_{1,2,1,1,1,2,1,2,1}	\\ 
& \qquad
-\Omega_{2,1,1,1,1,1,2,1,2}+\Omega_{2,1,1,1,1,1,2,2,1}+\Omega_{2,1,1,1,1,2,1,1,2}-\Omega_{2,1,1,1,1,2,1,2,1})
\bigr]\,,
\end{split}
\end{equation}
which has 20 massive fields, we encounter an interesting feature that has not appeared before. While usually, 
whenever we found no further fixed massless fields at a particular order, then this persisted up
until sextic terms in $W$. However, for this solution $G^{[12, 12]}_{(4)}$, we encounter 2 stabilized 
fields at cubic order and then for a particular component 0 stabilized fields at quartic order, followed 
again by 4 fixed fields at quintic order and 1 fixed field at sextic order in $W$. It shows explicitly 
that even if we encounter at a certain low order no further stabilization this could change again at higher order. 

The solution $G^{[12, 12]}_{(5)}$ below has 18 massive fields
\begin{equation}
\begin{split}
G^{[12, 12]}_{(5)}&=9 \bigl[ -\Omega_{1,1,1,1,2,1,2,1,2} + \Omega_{1,1,1,1,2,1,2,2,1} +
\Omega_{1,1,1,1,2,2,1,1,2} -\Omega_{1,1,1,1,2,2,1,2,1}	
\\ &\quad
+\omega(-\Omega_{1,1,1,2,1,1,2,1,2} + \Omega_{1,1,1,2,1,1,2,2,1} + \Omega_{1,1,1,2,1,2,1,1,2}
-\Omega_{1,1,1,2,1,2,1,2,1}	
\\ &\quad - \Omega_{1,1,2,1,1,1,2,1,2} +
\Omega_{1,1,2,1,1,1,2,2,1} + \Omega_{1,1,2,1,1,2,1,1,2} -\Omega_{1,1,2,1,1,2,1,2,1})\bigr]\,.
\end{split}
\end{equation}
We find that the quadratic equations resulting from cubic terms in $W$ give rise to two components. For one 
component we have four fixed fields and all higher order equations up to sextic terms in $W$ are then 
automatically solved in terms of the higher order terms in the 18 massive and 4 fixed fields. Another 
component has only 2 stabilized fields but this component is so complicated that we have not been able 
to solve higher order constraints, leading to the question marks in the table. This example exemplifies 
an interesting point. We would say that we have 18 massive and 2 stabilized massive fields at cubic order 
in $W$. It is in principle possible that more fields get stabilized if we were able to pursue the first 
component to quartic or quintic order. However, from the last row in the table we know that even when 
going to sextic power in the superpotential we cannot stabilized more than 4 massless fields in this model. 

We have also calculated cubic, quartic and quintic terms for the solution above that has the largest mass matrix rank to tadpole contribution $N_{\rm flux}$ and that is given above in \eqref{selfproclaimedreigningchampion}. Note this is not a physical solution since $N_{\rm flux}>12$. We find that there are no further stabilized fields up to quintic order in the superpotential. This is in line with the empirical observation from table \ref{tab:stabilization} that models with the largest mass matrix rank do not have fields that get stabilized at higher order in this $1^9$ model.

\section{Conclusion}\label{sec:conclusion} 

In this paper we have continued the study of an
orientifold of the $1^9$ Landau-Ginzburg model. The shortest vector problem in this model was
solved in \cite{Becker:2023rqi}. Specifically, it was shown that any (non-zero)
quantized $G_3$-flux in $H^{2,1}$ contributes at least $N_{\rm flux}=8$ to the tadpole 
cancellation condition: $N_{\rm flux}=12-N_{D3}$ . In a convenient basis one can write the 
flux as $G_3 = \sum_{\mathbf{l}} A^{\bf{l}} \Omega_{\mathbf{l}}$ and each non-zero flux component 
$A^{\bf{l}}$ will contribute at least 1 to $N_{\rm flux}$. Thus, any quantized flux configuration 
in this model can have at most 12 non-zero flux components \cite{Becker:2022hse}. An exhaustive 
search in \cite{Becker:2023rqi} proved that quantized flux solutions only exist for 4, 8 or more 
non-zero flux components. Furthermore, all solutions with 4 components were classified and a 
large class of 8 flux component solutions was presented. In this paper we have proven that this 
large class actually contains all 8 flux component solutions. These solutions are all related 
by an $S_7$ symmetry that is preserved after the orientifold projection so that there is essentially 
only one such solution
with 8 flux components. There are no known solutions with 9, 10 and 11 flux components but several
different ones with 12 flux components. It remains an important challenge for the future to fully
classify the flux configurations with 12 flux components and to prove the absence of solutions with
9, 10 or 11 components or to find such solutions. However, the full classification of all possible
flux configurations in this model seems now to be within reach.

Given the importance of moduli stabilization in trying to connect string theory to the real world,
4d $\mathcal{N}=1$ Minkowski vacua in this model were studied in \cite{Becker:2022hse}. 
It was found that all known solutions and some 
newly constructed ones have a large number of massless moduli. Out of the 64 complex scalar fields 
only between 14 and 26 were massive due to the
presence of the fluxes \cite{Becker:2022hse}. In this paper we have generated a large number of 
flux configuration with relatively small tadpole and calculated the number of massive fields. The 
scatter plot above in figure \ref{fig:tadvsmass} shows the tadpole contribution vs the the number 
of massive fields. We find a linear behavior as predicted by the tadpole conjecture but some of our solutions violate the refined version of the tadpole conjecture. For solutions within the tadpole bound $N_{\rm flux}\leq12$ we find $n_{\rm stab}/N_{\rm flux}=26/12 \sim 2.2 > 3/2$ and more generically we find a solution with $57/21 \sim 2.71$ that violates the refined tadpole conjecture by almost a factor of 2.

Lastly, we developed a procedure for systematically calculating higher order terms in the superpotential and checking whether there are massless fields that are stabilized in these Minkowski vacua. In addition to the eight different solutions discussed in \cite{Becker:2022hse}, we performed such a study of higher order stabilization for two more solutions from \cite{Becker:2023rqi}. In the latter paper it was found that the flux configuration with only four flux components come in three families with either 16, 22 or 26 massive fields. We have included one representative from all three of those, a representative from the single family with eight flux components discussed above and several solutions with twelve flux components. Thereby making this a relatively complete set of examples.

Our findings are summarized in table \ref{tab:stabilization} above and are interesting in many aspects. First, we actually find that some flux configurations do not stabilize massless fields via higher terms in the superpotential, even when including cubic, quartic, quintic and sextic terms. This might be due to the large symmetry group of this model and it would be interesting to understand better \cite{Grimm:2024fip}. For setups where higher order constraints appear, we are faced with solving polynomial equations in many variables and one might have expected that this is an insurmountable task. However, we actually found that these constraint equations are often solvable and they lead to the stabilization of several massless fields. We have found three example where massless fields get stabilized when including up to sextic terms in the superpotential. The total number of stabilized fields is then in all examples larger than the maximum number allowed by the refined version of the tadpole conjecture. Given that there is an infinite number of higher order terms in the superpotential it is not clear whether and how many more moduli will be stabilized at even higher order. We are currently at the limit of what can be calculated with a normal computer and it would be interesting to use more powerful computers or to develop more sophisticated techniques to extend our result to higher order. We leave this as an exciting challenge for the future.

\section*{Acknowledgments}
We would like to thank James Gray and Daniel Junghans for useful discussions, and Mariana Gra\~na for valuable feedback on an initial draft. The work of KB and AS is supported in part by the 
NSF grant PHY-2112859. MR acknowledges the support of the Dr. Hyo Sang Lee Graduate Fellowship from the 
College of Arts and Sciences at Lehigh University. The work of MR and TW is supported in part by the NSF 
grant PHY-2210271. This research was supported in part by grant NSF PHY-2309135 to the Kavli Institute 
for Theoretical Physics (KITP). JW thanks the International Centre for Mathematical Sciences, Edinburgh, 
for support and hospitality during the ICMS Visiting Fellows programme where this work was completed. This 
work was supported by EPSRC grant EP/V521905/1. This work is funded by the Deutsche Forschungsgemeinschaft 
(DFG, German Research Foundation) under Germany’s Excellence Strategy EXC 2181/1 — 390900948 (the Heidelberg 
STRUCTURES Excellence Cluster).

\appendix



\section{Proof that all \texorpdfstring{$t^{\kk}$}{t*k} appear in
\texorpdfstring{$W$}{W}}\label{app:allmoduli} Given that it is difficult to stabilize all fields,
one might ask whether one can show that flat directions arise due to the simple fact that some
fields do not appear in the superpotential at all. This is however not the case and in this
appendix, we prove that the dilaton and all 63 complex structure moduli do appear (at higher order)
in $W$ for any non-zero flux choice.

Let us assume that we turn on some flux and generate thereby a mass term for some fields, i.e., we
assume that $\partial_{t^{\mathbf{k}_1}}\partial_{t^{\mathbf{k}_2}} W \neq 0$ for some
$\mathbf{k}_1, \mathbf{k}_2$. This is true for any non-zero flux choice and implies from equation
\eqref{eq:L} above that 
\begin{equation}
    \mathbf{\bar{l}}=(\mathbf{k}_1+\mathbf{k}_2 +{\bf 1})\bmod 3 \,.
\end{equation} 
Now we ask whether a $t^{\mathbf{k}}$ exists that does not appear in $W$. The answer is no as can be
seen as follows: Since $\mathbf{k}$ contains only 0's and 1's we have that $3\cdot \mathbf{k}=0 \mod
3$. So at quintic order in the superpotential, there is a term proportional to $t^{\mathbf{k}_1}
t^{\mathbf{k}_2} (t^{\mathbf{k}})^3$ since we have
\begin{equation}
     \mathbf{\bar{l}}=(\mathbf{k}_1+\mathbf{k}_2 +3\cdot \mathbf{k} + {\mathbf 1}) \bmod 3 = (\mathbf{k}_1+\mathbf{k}_2 + {\mathbf{1}}) \bmod 3\,.  
\end{equation}
Thus at quintic order, every $t^{\mathbf{k}}$ will appear for sure. However, it will do so in a
rather simple way multiplied by terms that already appeared at quadratic order and thus these terms
cannot really stabilize $t^{\mathbf{k}}$. In concrete examples, we usually find that all fields
already appear when including quartic terms in $W$.

\bibliographystyle{JHEP}
\bibliography{refs}

\end{document}